\documentclass[12pt]{article}
\usepackage{amsfonts}
\usepackage{amsmath, amssymb}
\usepackage{youngtab}
\usepackage{hyperref}
\usepackage[dvips]{color}
\usepackage{psfrag}
\usepackage{feynmp}
\usepackage[dvipdfmx]{graphicx}
\usepackage{braket}
\usepackage{bm}
\usepackage{subfigure}

\usepackage{comment}
\includecomment{pdffig}

\allowdisplaybreaks[1]

\Yboxdim{5pt}

\makeatletter
    
    \@addtoreset{equation}{section}
  \makeatother

\usepackage{color}

\def\rem#1{}

\renewcommand{\title}[1]{\vbox{\center\LARGE{#1}}\vspace{5mm}}
\renewcommand{\author}[1]{\vbox{\center\large#1}\vspace{5mm}}

\begin{document}
\bibliographystyle{utphys}

\begin{titlepage}
\begin{center}
\vspace{5mm}
\hfill {\tt 
}\\
\vspace{20mm}

\title{  
\LARGE  Euler transformation for multiple $q$-hypergeometric series from  wall-crossing formula of $K$-theoretic vortex partition function 
}

\vspace{7mm}
{
Yutaka Yoshida\footnote{yutakayy@law.meijigakuin.ac.jp} 
\\
}

\vspace{3mm}
{\small {\it Department of Current Legal Study, Faculty of Law, Meiji Gakuin University, 1-2-37 
Shirokanedai, Minato-ku, Tokyo 108-8636, Japan}} \\
{\small {\it Institute for Mathematical Informatics, Meiji Gakuin University
1518 Kamikurata-cho, Totsuka-ku, Yokohama 244-8539, Japan}}
\end{center}

\vspace{7mm}
\abstract{
We show that transformation formulas for multiple $q$-hypergeometric series agree
with wall-crossing formulas for $K$-theoretic vortex partition functions obtained by
Hwang, Yi, and the author \cite{Hwang:2017kmk}.
For vortex partition functions in 3d $\mathcal{N}=2$ gauge theories, we show that
the wall-crossing formula agrees with the Kajihara transformation
\cite{kajihara2004euler}.
For vortex partition functions in 3d $\mathcal{N}=4$ gauge theories, we show that
the wall-crossing formula agrees with the transformation formula of Halln\"as,
Langmann, Noumi, and Rosengren \cite{Halln_s_2022}.
Since the $K$-theoretic vortex partition functions are related to indices such as
the $\chi_t$-genus of the handsaw quiver variety, we discuss a geometric
interpretation of Euler transformations in terms of wall-crossing formulas for the
handsaw quiver variety.
}
\vfill

\end{titlepage}

\tableofcontents

\section{Introduction and summary}
Dualities relate supersymmetric quantum field theories, which typically have the same fixed
point in the renormalization group flow.
At the level of supersymmetric indices and also partition functions, a duality implies
the agreement of the supersymmetric indices of the dual pair.
Combined with the supersymmetric localization formula, the agreement of the indices
predicts many non-trivial identities between special functions.
For example, in four dimensions, the equality of superconformal indices
\cite{Romelsberger:2005eg, Kinney:2005ej} of a dual pair predicts a non-trivial
identity between integrals of elliptic gamma functions.

In three dimensions, supersymmetric localization formulas for supersymmetric indices
and partition functions
\cite{Kim:2009wb, Hama:2011ea, Imamura:2011su, Benini:2015noa} for $U(N)$ gauge
theories on a closed three-manifold $M_3$ admit a remarkable factorization into
a product of a pair of $K$-theoretic vortex partition functions
\cite{Pasquetti:2011fj, Hwang:2012jh, Beem:2012mb, Taki:2013opa, Fujitsuka:2013fga, Benini:2013yva}:
\begin{align}
Z_{M_3}= \sum Z_{1\text{-loop}} Z_{\rm vortex} Z_{\rm vortex},  \qquad (M_3=S^3_b, S^1 \times S^2)\,.
\label{eq:partition1}
\end{align}
Here, $Z_{\rm vortex}$, called {\it the $K$-theoretic vortex partition function}, is
the generating function for Witten indices of supersymmetric quantum mechanics (SQM)
associated with the handsaw quiver of type $A_1$.
Since the moduli space of vortices given by the Higgs branch vacua of SQM is
isomorphic to a handsaw quiver variety of type $A_1$, the $K$-theoretic vortex
partition function agrees with the generating function for an equivariant index such
as the $\chi_t$-genus of handsaw quiver varieties of type $A_1$.

The problem of finding the precise relation among the $Z_{M_3}$'s for a
three-dimensional (3d) Seiberg-like dual pair reduces to finding the relation
between the $K$-theoretic vortex partition functions of the Seiberg-like dual
pair \cite{Hwang:2015wna}. Later, it was found \cite{Hwang:2017kmk} that the
relation between the $K$-theoretic vortex partition functions of a Seiberg-like
dual pair agrees with the wall-crossing (WC) formula of the $K$-theoretic vortex
partition function with respect to variation of the Fayet-Iliopoulos (FI)
parameter of SQM.

In this article, we show that the wall-crossing formulas for vortex partition
functions in 3d $\mathcal{N}=2$ and $\mathcal{N}=4$ $U(N)$ gauge theories obtained
in \cite{Hwang:2015wna, Hwang:2017kmk} agree with Euler transformations of
$q$-hypergeometric series studied by Kajihara \cite{kajihara2004euler} and by
Halln\"as, Langmann, Noumi, and Rosengren \cite{Halln_s_2022}, respectively.
These Euler transformations can be regarded as generalizations of the Euler
transformation for the Gauss hypergeometric series:
\begin{align}
{}_2F_1(a,b;c;z) = (1-z)^{c-a-b} {}_2F_1(c-a,c-b;c;z)\,.
\label{eq:Gauss}
\end{align}
 Since the $K$-theoretic vortex partition functions are indices for handsaw quiver
varieties, the wall-crossing formulas give geometric interpretations of Euler
transformations of $q$-hypergeometric series.
  
This article is organized as follows.
In Section \ref{section:Kvortex}, we review the $K$-theoretic vortex partition
functions in 3d $\mathcal{N}=2$ Chern-Simons matter theory (a gauge theory with a
Chern-Simons term) and explain the relation between an ADHM-like description of the
moduli space of vortices and a handsaw quiver variety of type $A_1$.
We then obtain two distinct expressions for the supersymmetric localization formula
in the regions of positive and negative 1d FI parameter, respectively.
In Section \ref{sec:wallcross}, following \cite{Hwang:2017kmk}, we prove the
wall-crossing formula, which relates the vortex partition function in the positive
FI-parameter region to that in the negative FI-parameter region.
In Section \ref{sec:Kajihara}, we give the precise parameter identification between
the vortex partition function and the Kajihara transformation.

We perform a similar computation for a 3d $\mathcal{N}=4$ gauge theory in Section
\ref{sec:HLNRformula}.
In Section \ref{sec:vortex3dN4}, we review the $K$-theoretic vortex partition
function for the 3d $\mathcal{N}=4$ theory.
In Section \ref{sec:WCHLNR}, we show that the wall-crossing formula of the vortex
partition function agrees with a transformation formula of trigonometric type
\cite{Halln_s_2022}.
Next, we explain the geometric interpretation of the vortex partition functions in
Section \ref{sec:chitgenus}.
For 3d $\mathcal{N}=4$ gauge theories, we find that the $K$-theoretic vortex
partition function is equal to the equivariant $\chi_t$-genus up to a trivial factor.
For 3d $\mathcal{N}=2$ gauge theories, although the $K$-theoretic $k$-vortex
partition function and the corresponding index are related up to a factor, the
generating functions are different.
We briefly comment on an elliptic analogue of the wall-crossing formula in Section
\ref{sec:elliptic}.
 In Section \ref{sec:2dlimit}, we perform the dimensional reduction from three
dimensions to two dimensions and derive the wall-crossing formulas for vortex
partition functions in 2d $\mathcal{N}=(2,2)$ and $\mathcal{N}=(2,2)^*$ gauge
theories.
In Section \ref{sec:future}, we comment on future directions of this work.

\section{Kajihara transformation as wall-crossing formula}
\label{eq:vortex}

In this section, we show that the wall-crossing formula of the $K$-theoretic
vortex partition function derived in \cite{Hwang:2017kmk} agrees with the
Kajihara transformation of multiple $q$-hypergeometric series
\cite{kajihara2004euler}.

\subsection{$K$-theoretic vortex partition function and handsaw quiver variety}
\label{section:Kvortex}

\begin{figure}[t]
  \centering
  \includegraphics[height=5cm]{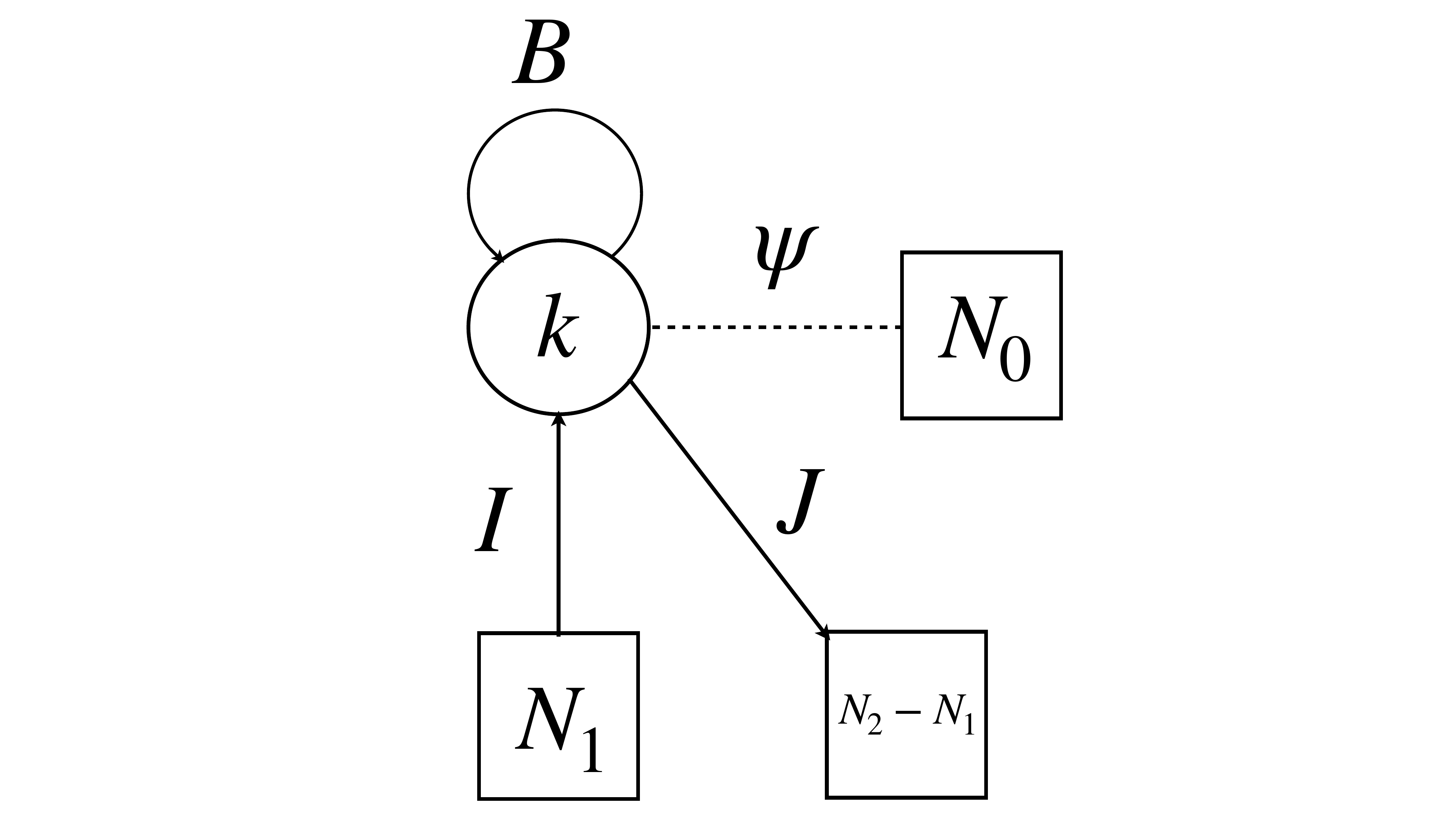}
  \caption{The quiver diagram representing the 1d $\mathcal{N}=2$ SQM associated
  with the ADHM-like description of the $k$-vortex moduli space for the 3d
  $\mathcal{N}=2$ gauge theory in Table \ref{table:charge1}.
  The circle labeled $k$ denotes the 1d $\mathcal{N}=2$ $U(k)$ vector multiplet.
  A solid arrow denotes a 1d $\mathcal{N}=2$ chiral multiplet. A dashed line
  denotes the 1d $\mathcal{N}=2$ Fermi multiplet. $I$ and $J$ denote the
  scalar fields in the chiral multiplets. $\psi$ denotes the fermion in the
  Fermi multiplet. The representations of the multiplets are summarized in Table
  \ref{table:HScharge}}
  \label{fig:handsaw1}
\end{figure}

First, we review the vortex partition function of a 3d $\mathcal{N}=2$
Chern-Simons matter theory with unitary gauge group $U(N_1)$ coupled to $N_2$
chiral multiplets in the fundamental representation of $U(N_1)$ and $N_0$
chiral multiplets in the anti-fundamental representation of $U(N_1)$.
In this paper, we assume $N_0 \le N_2$.
The precise matter content of the 3d gauge theory is summarized in Table
\ref{table:charge1}.
As we will show, the case with $N_2=N_0$ and zero Chern-Simons level,
$\kappa=0$, is relevant to the Kajihara transformation.

Let us consider the partition function (supersymmetric index) of the 3d gauge
theory on $S^1 \times \mathbb{R}^2$ around solutions to the vortex equations:
\begin{align}
&F_{12}=\frac{g^2}{2}(\phi \phi^{\dagger} -\xi 1_{N_1 \times N_1}), \nonumber \\
&(D_{1}+{\rm i} D_2) \phi=0\,.
\label{eq:vortexeq}
\end{align}
with $k$-vortex number:
$k= \frac{1}{2 \pi}  \int_{\mathbb{R}^2} \mathrm{Tr} F_{12}$.
Here $F_{\mu \nu}=\partial_{\mu} A_{\nu}-\partial_{\nu} A_{\mu} +{\rm i} [A_{\mu}, A_{\nu}]$
is the field strength of the $U(N_1)$ gauge field $\{ A_{\mu} \}_{\mu=0}^2$.
$D_{\mu}=\partial_{\mu}+{\rm i}A_{\mu}$ is the covariant derivative.
As explained in Table \ref{table:charge1}, $\phi$ denotes the $N_2$-tuple of
scalars in the fundamental representation of the gauge group $U(N_1)$.
$\xi$ is the FI parameter for the 3d gauge theory.
Since the moduli space of solutions to the vortex equations admits an ADHM-like
description in terms of the D-brane construction \cite{Hanany:2003hp} and the
field-theoretic construction \cite{Eto:2005yh, Eto:2006pg},
the partition function of the 3d gauge theory around the $k$-vortex background on
$S^1 \times \mathbb{R}^2$ reduces to the Witten index of 1d $\mathcal{N}=2$ SQM
on $S^1$ associated with the handsaw quiver of type $A_1$ depicted in Figure
\ref{fig:handsaw1}. The matter content of the SQM is summarized in Table
\ref{table:HScharge}.

In the SQM, the moduli space of Higgs branch vacua is given by a K\"ahler
quotient, which is isomorphic to a handsaw quiver variety of type $A_1$:
\begin{align}
\mathcal{M}^{ (\zeta)}_{k,N_1,N_2}&= \Bigl\{ (B,I,J) \Big| [B, B^{\dagger}] +I I^{\dagger}- J^{\dagger} J =\zeta 1_{k \times k}  \Bigr\}\Big/U(k) \,,
\label{eq:Higgsvac1} \\
&\simeq \{ (B,I,J)\}/GL(k,\mathbb{C})\,.
\label{eq:Higgsvac1GL}
\end{align}
Here
\begin{align}
B \in \mathrm{End}(V), \,\,
I \in \mathrm{Hom} (W^{(1)}, \,\, V), \,\, J \in \mathrm{Hom} (V, W^{(2)})
\end{align}
with $V \simeq \mathbb{C}^k$, $W^{(1)} \simeq \mathbb{C}^{N_1}$, and
$W^{(2)} \simeq \mathbb{C}^{N_2-N_1}$.
These are the vacuum expectation values of the scalars in the 1d
$\mathcal{N}=2$ chiral multiplets.
In \eqref{eq:Higgsvac1} and \eqref{eq:Higgsvac1GL}, $g \in U(k)$ or
$GL(k,\mathbb{C})$ acts on $(B, I,J)$ as
$(B, I,J) \mapsto (g B g^{-1}, g I, J g^{-1})$.
The quotients in \eqref{eq:Higgsvac1} and \eqref{eq:Higgsvac1GL} represent the
K\"ahler quotient and the quotient with a stability condition.

The positive and negative FI-parameter regions in \eqref{eq:Higgsvac1} correspond
to two different stability conditions of the handsaw quiver variety, respectively.
According to the D-brane construction \cite{Hanany:2003hp} and the field-theoretic
construction \cite{Eto:2005yh, Eto:2006pg} of the vortex moduli, the moduli space
of Higgs branch vacua $\mathcal{M}^{ (\zeta>0)}_{k,N_1,N_2}$ is isomorphic to the
moduli space of $k$-vortex solutions of \eqref{eq:vortexeq}.

The Witten index for the 1d $\mathcal{N}=2$ SQM depicted by the handsaw quiver of
type $A_1$ is called the {$K$-theoretic $k$-vortex partition function} for the 3d
$\mathcal{N}=2$ gauge theory.
Since the Witten index is related to an (equivariant) index of the moduli space of
Higgs branch vacua of the SQM, the $K$-theoretic vortex partition functions give
indices of the handsaw quiver variety.
Note that the fermion $\psi$ in the Fermi multiplet corresponds to a characteristic
class integrated over the handsaw quiver variety.
This situation is analogous to the instanton case: the Witten indices for SQM
depicted by the ADHM (Jordan) quiver are called $K$-theoretic instanton partition
functions \cite{Nekrasov:2004vw, Hwang:2014uwa}, which give indices of instanton
moduli spaces.
We will explain the geometric interpretation of the vortex partition functions in
Section \ref{sec:chitgenus}.

\begin{table}[thb]
\begin{center}
\begin{tabular}{c || c c c   }
			&	$U(N_0)$ &	$U(N_1)$ & $U(N_2)$  
			 	\\
			 \hline
$ \phi$	 &	${\bm 1}$	&	${\bm N_1}$	& $\overline{\bm N_2}$ 	\\
$\tilde{\phi}$	 &	${\bm N_0}$	&	$\overline{\bm N_1}$	& ${\bm 1}$ 	\\
\end{tabular} 
\caption{The representations of the scalars in the 3d $\mathcal{N}=2$ chiral
multiplets. $G=U(N_1)$ is the gauge group of the 3d theory.
$G_F=(U(N_0) \times U(N_2))/U(1)$ is the flavor symmetry group.
Here, $\phi \in \mathrm{Hom}(\mathbb{C}^{N_1}, \mathbb{C}^{N_2})$ and
$\tilde{\phi} \in \mathrm{Hom}(\mathbb{C}^{N_0}, \mathbb{C}^{N_1})$ denote the
scalars in the chiral multiplets.
${\bm N_s}$ (resp. $\overline{\bm N_s}$) denotes the fundamental
(resp. anti-fundamental) representation of $U(N_s)$.
${\bm 1}$ denotes the trivial representation.
}
\label{table:charge1}
\end{center}
\end{table}

\begin{table}[thb]
\begin{center}
\begin{tabular}{c || c c c  c c }
			&	$U(k)$ &	$U(N_1)$ & $U(N_2-N_1)$ & $U(N_0)$ & $U(1)_{\varepsilon}$
			 	\\
			 \hline
$ B$	 &	$\bm{ adj}$	&	${\bm 1}$	& ${\bm 1}$ & ${\bm 1}$  & $-1$	\\
$ I$	 &	${\bm k}$	&	$\overline{\bm N_1}$	& ${\bm 1}$ & ${\bm 1}$ & $-\frac{1}{2}$\\
$ J$	 &	$\bar{\bm k}$	&	${\bm 1}$	& ${{\bm N_2}-{\bm N_1}}$ & ${\bm 1}$ & $-\frac{1}{2}$	\\
$ \psi$	 &	$\bar{\bm k}$	&	${\bm 1}$	& ${\bm 1}$ & ${\bm N_0}$ & $-\frac{1}{2}$	\\
\end{tabular} 
\caption{The representations of the scalars and fermions in the 1d
$\mathcal{N}=2$ multiplets describing the $k$-vortex moduli space in the 3d
$\mathcal{N}=2$ $U(N_1)$ gauge theory depicted by the quiver diagram in Figure
\ref{fig:handsaw1}. The fields $B$, $I$, and $J$ are scalars in the 1d
$\mathcal{N}=2$ chiral multiplets. The field $\psi$ is the fermion in the 1d
$\mathcal{N}=2$ Fermi multiplet. $U(k)$ is the gauge group. $U(N_1)$,
$U(N_2-N_1)$, $U(N_0)$, and $U(1)_{\varepsilon}$ are flavor symmetry groups.
'$\bm{adj}$' denotes the adjoint representation.
}
\label{table:HScharge}
\end{center}
\end{table}


The $K$-theoretic $k$-vortex partition function $Z^{(\zeta)}_{k}$ can be
evaluated by applying the supersymmetric localization formula for the Witten
index \cite{Hwang:2014uwa, Hori:2014tda}.
This yields the following residue integral expression:
\begin{align}
Z^{(\zeta)}_{k}&=
\frac{1}{k! (2 \sinh \frac{-\varepsilon}{2})^k } \oint_{JK(\zeta)} \prod_{\alpha=1}^k \frac{d u_{\alpha}}{2 \pi {\rm i}} 
\prod_{ 1 \le \alpha \neq  \beta \le k } \frac{2 \sinh \frac{u_{\alpha}-u_{\beta}}{2}  } {2 \sinh \frac{u_\alpha-u_{\beta}-\varepsilon}{2} } 
\nonumber \\
& \qquad  \times 
 \prod_{\alpha=1}^{k} \frac{e^{\kappa u_\alpha} \prod_{j=1}^{N_{0}} 2 \sinh \left( {\frac{u_\alpha-\tilde{m}_j-\varepsilon/2}{2}} \right) }
 {\prod_{j=1}^{N_1} 2 \sinh \left( \frac{u_\alpha-m_j -\varepsilon/2 }{2} \right) \cdot \prod_{j =N_1+1}^{N_2} 2 \sinh \left( \frac{-u_\alpha+m_j-\varepsilon/2}{2} \right)} \,.
 \label{eq:contour1}
\end{align}
In \eqref{eq:contour1}, the complex parameters $\{ m_i \}_{i =1}^{N_1}$,
$\{ m_i \}_{i=N_1+1}^{N_2}$, and $\{ \tilde{m}_i \}_{i=1}^{N_0}$ are the
chemical potentials for the 1d flavor symmetry groups $U(N_1)$, $U(N_2-N_1)$,
and $U(N_0)$ acting on $I$, $J$, and $\psi$, respectively.
The chemical potential $\varepsilon$ is called the $\Omega$-background
parameter.

Later, $e^{m_i}$ and $e^{\varepsilon}$ are identified with the
$(\mathbb{C}^*)^{N_2+1}$-equivariant parameters associated with the
$U(N_1) \times U(N_2-N_1) \times U(1)_{\varepsilon}$ action on
$\mathcal{M}^{(\zeta)}_{k, N_1,N_2}$ defined by
\begin{align}
 & (B,I,J) \mapsto (B,I g^{-1}_1,J), \quad g_1 \in U(N_1), \nonumber \\
& (B,I,J) \mapsto (B,I,g_2 J), \quad g_2 \in U(N_2-N_1),
\label{eq:Gaction}\\
& (B,I,J) \mapsto (g_3 B, g_3 I, g_3 J), \quad g_3 \in U(1)_{\varepsilon} \,. \nonumber
\end{align}

$\kappa \in \mathbb{Z}$ or $\mathbb{Z}+\frac{1}{2}$ is the 1d Chern-Simons level
induced by the 3d supersymmetric Chern-Simons term in the $U(N_1)$ gauge theory:
\begin{align}
{S}_{ CS}&=\frac{\kappa}{4 \pi {\rm i}} \int_{S^1 \times \mathbb{R}^2}  \mathrm{Tr} \left(A \wedge d A-\frac{2 {\rm i}}{3} A^3 \right)+\cdots \,.
\label{eq:3dCSterm}
\end{align}
The trace in \eqref{eq:3dCSterm} is taken in the fundamental representation of
$\mathfrak{u}(N_1)$.
In this article, we assume
$-\frac{N_2-N_0}{2} \le \kappa \le \frac{N_2-N_0}{2}$ and
$\kappa+\frac{N_2-N_0}{2} \in \mathbb{Z}$.

Let us evaluate the contour integral in \eqref{eq:contour1}.
The contour is determined by the Jeffrey-Kirwan (JK) residues, which are
correlated with the sign of the 1d FI parameter $\zeta$ in the SQM
\cite{Hori:2014tda}.
The JK residues are evaluated as follows \cite{Hwang:2017kmk}.

When the 1d FI parameter is positive, $\zeta > 0$, the JK residues in
\eqref{eq:contour1} are evaluated at the poles defined by
\begin{align}
u_{\alpha}-u_{\beta} -\varepsilon&=0, \quad  \alpha, \beta  \in \{ 1,\cdots, k \}, \nonumber \\
u_{\alpha} -m_i -\frac{\varepsilon}{2}&=0, \quad  \alpha \in \{1,\cdots, k\}, \, \,i\in \{1,\cdots, N_1 \}.
\label{eq:hyperplanes1}
\end{align}
The intersection points of the hyperplanes in \eqref{eq:hyperplanes1} are
classified by
\begin{align}
u_{\alpha=\ell_i}=m_{i} +(\ell_i-1/2)\varepsilon, \quad \ell_i =1,\cdots, k_i,  \,\,\,i \in \{1,\cdots, N_1 \},
\label{eq:solu1}
\end{align}
where $k_i \ge 0$ and $\sum_{i =1}^{N_1} k_i=k$.

On the other hand, the JK residues in the negative FI-parameter region,
$\zeta < 0$, are evaluated at the poles given by
\begin{align}
u_{\alpha} -u_{\beta} -\varepsilon&=0, \quad  \alpha, \beta \in \{1,\cdots, k\}, \nonumber \\
-u_{\alpha}+m_j -\frac{\varepsilon}{2}&=0, \quad  \alpha \in \{ 1,\cdots, k\}, \, \,j \in \{N_1+1,\cdots, N_2\}.
\label{eq:hyperplanes2}
\end{align}
The intersection points of the hyperplanes in \eqref{eq:hyperplanes2} are
classified by
\begin{align}
u_{\alpha=\ell_i}=m_{i} -(\ell_i-1/2)\varepsilon, \quad \ell_i =1,\cdots, k_i,  \,\,\,i \in \{N_1+1,\cdots, N_2 \},
\label{eq:solu2}
\end{align}
where $k_i \ge 0$ and $ \sum_{i =N_1+1}^{N_2} k_i=k$.
Then the $K$-theoretic $k$-vortex partition functions in the positive and negative
FI-parameter regions are given by
{\small
\begin{align}
 Z^{(\zeta>0)}_{k}&=\sum_{k_1+\cdots +k_{N_1}=k} \frac{ (-1)^{N_1k} e^{\kappa \sum_{i =1}^{N_1} (k_i m_i+ k^2_i \varepsilon/2) }\prod_{i =1}^{N_1} \prod_{j=1}^{N_{0}}   \prod_{\ell_i=1}^{k_i}  2 \sinh \left({\frac{m_i-\tilde{m}_j+ (\ell_i-1) \varepsilon}{2}} \right)}
 {\prod_{i,j =1}^{N_1}  \prod_{\ell_i=1}^{k_i} 2 \sinh \left( \frac{m_j-m_i -(\ell_i -k_j-1)\varepsilon }{2} \right)  \prod_{i =1}^{N_1} \prod_{j =N_1+1}^{N_2} \prod_{\ell_i=1}^{k_i} 2 \sinh \left( \frac{-m_i+m_j-\ell_i\varepsilon}{2} \right)}, 
 \label{eq:kvortexplus}
  \\
 Z^{(\zeta<0)}_k
 &=\sum_{ {k}_{N_1+1}+\cdots +{k}_{N_2}={k}}  \frac{(-1)^{(N_2-N_1+N_0)k}  e^{-\kappa \sum_{i =N_1+1}^{N_2} (-k_i m_i+k^2_i \varepsilon/2)} \prod_{i =N_1+1}^{N_2} \prod_{j=1}^{N_{0}}   \prod_{\ell_i=1}^{k_i}  2 \sinh \left({\frac{-m_i+\tilde{m}_j+ \ell_i \varepsilon}{2}} \right)}
 {\prod_{i, j =N_1+1}^{N_2}  \prod_{\ell_i=1}^{k_i} 2 \sinh \left( \frac{-m_j+m_i-(\ell_i -k_j-1)\varepsilon }{2} \right)  \prod_{i =N_1+1}^{N_2} \prod_{j =1}^{N_1} \prod_{\ell_i=1}^{k_i} 2 \sinh \left( \frac{m_i-m_j-\ell_i\varepsilon}{2} \right)} .
  \label{eq:kvortexminus}
 \end{align}
}

Since the sign of the FI parameter corresponds to the two distinct stability
conditions that define the handsaw quiver variety, the supersymmetric
localization formula for the $K$-theoretic $k$-vortex partition functions
\eqref{eq:kvortexplus} and \eqref{eq:kvortexminus} should be related to indices
of handsaw quiver varieties with two different stability conditions.
The vortex partition functions in the positive and negative FI-parameter regions
may have different values: $Z^{(\zeta>0)}_k \neq Z^{(\zeta<0)}_k$, and such a
situation is called a wall-crossing phenomenon.

\subsection{Derivation of the wall-crossing formula for the vortex partition function}
\label{sec:wallcross}

Following \cite{Hwang:2017kmk}, we derive the relation between
$Z^{(\zeta>0)}_k$ and $Z^{(\zeta<0)}_k$, i.e., the wall-crossing formula for the
$K$-theoretic vortex partition function, which is closely related to the
wall-crossing formula for indices of handsaw quiver varieties of type $A_1$.

To derive the wall-crossing formula, we first introduce a multiple contour
integral defined by
\begin{align}
\mathcal{Z}_{k}
&:= \oint_{T^k_{x}}   \frac{d x_{k}}{2 \pi {\rm i} x_{k}}  \cdots \frac{d x_{2}}{2 \pi {\rm i} x_{2}} \frac{d x_{1}}{2 \pi {\rm i} x_{1}}  \mathcal{I}_k(x_1,x_2,\cdots, x_k)\,,
 \label{eq:tildeZk}
\end{align}
where
\begin{align}
\mathcal{I}_k(x_1,\cdots,x_{k})&=\frac{q^{\frac{k^2}{2}} }{k! (1-q)^k }  (\prod_{i=1}^{N_2} x^{(2)}_i)^{\frac{k}{2}} (\prod_{i=1}^{N_0} x^{(0)}_i)^{-\frac{k}{2}}  q^{\frac{(2N_1-N_2-N_0) k}{4} }
\prod_{1 \le \alpha \neq  \beta \le k}  
\frac{x_{\alpha}-x_{\beta} } {x_{\alpha}-q x_{\beta} } \nonumber \\
& \times \prod_{\alpha=1}^{k} \frac{ x^{\kappa+\frac{N_2-N_0}{2}}_{\alpha} \prod_{j=1}^{N_{0}} ( x_{\alpha}- x^{(0)}_j q^{\frac{1}{2}}  ) }
 {\prod_{i =1}^{N_1} ( x_{\alpha}- x^{(2)}_i q^{\frac{1}{2}}  ) \cdot \prod_{j =N_1+1}^{N_2} (  x^{(2)}_j q^{-\frac{1}{2}} -x_{\alpha})} \,.
 \label{eq:integrand2}
\end{align}
Here, ${T}^k_x$ is the $k$-dimensional torus defined by
${T}^k_{x}=\{ (x_1,\cdots,x_k) \in \mathbb{C}^k \mid |x_{1}|=\cdots=|x_{k}|=1\}$.
We assume that the variables $q, x_i^{(2)} \in \mathbb{C}$ lie in the following
regions:
\begin{align}
| q| <1, \,\,  |x^{(2)}_{i} q^{n+\frac{1}{2}}| <1,  \,\,|x^{(2)}_{j} q^{-n-\frac{1}{2}}| >1,
\end{align}
for $n=0,1,2,\cdots$, $i=1,2,\cdots,N_1$, and $j=N_1+1,\cdots,N_2$.

The multi-contour integral in \eqref{eq:tildeZk} is evaluated either by the
residues inside the torus ${T}^k_x$, or by the residues outside the torus.
First, we consider evaluating the residues inside the torus. In this case,
the sequences of poles that contribute to the iterated residues are divided
into the following two types.
The first type includes at least one pole at the origin, $x_i=0$, of the form
\begin{align}
&\{ x_{i_1}= q x_{i_2}, \, x_{i_2} = q x_{i_3},\, \cdots, x_{i_{n-1}}= q x_{i_n}, \,x_{i_n} =0 \}\,,\nonumber \\
&\{ x_{i_1}= q x_{i_3}, \, x_{i_2} = q^2 x_{i_3}, \,\cdots, x_{i_{n-1}}= q x_{i_n}, \,x_{i_n} = 0\}\,, \nonumber \\
&\quad \vdots \nonumber \\
&\{ x_{i_1}= 0, \, x_{i_2} = 0, \cdots, x_{i_{n-1}}= 0, \,x_{i_n} = 0\}\,.
\label{eq:poleszero}
\end{align}
Here $1 \le i_1 < i_2 < \cdots < i_n \le k$ and $0 \le n \le k$.
The second type is given by sequences of poles that do not include the pole at
the origin:
\begin{align}
&\{ x_{i_{n+1}}= q x_{i_{n+2}}, \, x_{i_{n+2}} = q x_{i_{n+3}},\, \cdots, x_{i_{k-1}}= q x_{i_k}, \,x_{i_k} = x^{(2)}_{j_1} q^{\frac{1}{2}} \},\nonumber \\
&\{ x_{i_{n+1}}= q x_{i_{n+3}}, \, x_{i_{n+2}} = q^2 x_{i_{n+3}}, \,\cdots, x_{i_{k-1}}= q x_{i_k}, \,x_{i_k} = x^{(2)}_{j_1} q^{\frac{1}{2}} \}, \nonumber \\
&\quad \vdots \nonumber \\
&\{ x_{i_{n+1}}= x^{(2)}_{j_1} q^{\frac{1}{2}}, \, x_{i_{n+2}} = x^{(2)}_{j_2} q^{\frac{1}{2}}, \cdots, x_{i_{k-1}}=x^{(2)}_{j_{k-n-1}} q^{\frac{1}{2}} , \,x_{i_k} = x^{(2)}_{j_{k-n}} q^{\frac{1}{2}} \}.
\label{eq:polesmass}
\end{align}
Here
$j_1, \cdots, j_{k-n} \in \{1,\cdots, N_1 \}$.
Except in the case where $\kappa=\frac{N_2-N_0}{2}$, the residues at the poles
of the first type vanish.
In this case,
$\mathcal{Z}_k$ is given by the residues of the second type, which agrees with
the $k$-vortex partition function \eqref{eq:kvortexplus}.
This is because $\mathcal{I}_k(x_1,\cdots,x_k)$ is the same as the integrand of
\eqref{eq:contour1} under the following identification of the parameters
\begin{align}
x_{\alpha}&=e^{u_{\alpha}}, \,\, 
 x^{(0)}_a =e^{\tilde{m}_a}, \,\, x^{(2)}_a =e^{m_a}, \,\, q=e^{\varepsilon},
 \label{eq:idenpara}
\end{align}
and also because the poles of the second type are the same as the poles
\eqref{eq:hyperplanes1} that appear in the JK residue operation under the
identification \eqref{eq:idenpara}.

On the other hand, when the condition $\kappa = \frac{N_2-N_0}{2}$ is satisfied,
the residues at poles of the first type \eqref{eq:poleszero} are non-zero.
The sum of the residues of the first type with respect to
$x_{i_1},\cdots, x_{i_n}$ is evaluated as follows.
Since the second line in \eqref{eq:integrand2} is regular at the origin of
$(x_{i_1},\cdots,x_{i_n})$, we may replace this factor in the residue
computation by
\begin{align}
&
\prod_{ \alpha \in \mathcal{J}}   \prod_{ \beta \in \mathcal{J}^c}
\frac{x_{\alpha}-x_{\beta} } {x_{\alpha}-q x_{\beta} }  \frac{x_{\beta}-x_{\alpha} } {x_{\beta}-q x_{\alpha} }
 \cdot
 \prod_{\alpha \in \mathcal{J}} \frac{ x^{\kappa+\frac{N_2-N_0}{2}}_{\alpha} \prod_{j=1}^{N_{0}} ( x_{\alpha}- x^{(0)}_j q^{\frac{1}{2}}  ) }
 {\prod_{i =1}^{N_1} ( x_{\alpha}- x^{(2)}_i q^{\frac{1}{2}}  ) \cdot \prod_{j =N_1+1}^{N_2} (  x^{(2)}_j q^{-\frac{1}{2}} -x_{\alpha})}
 \nonumber \\
 &\underset{ \{  x_{i}=0 \}_{i \in \mathcal{J}} }{\longrightarrow}   (-1)^{N_0-N_1} q^{-2n(k-n)+(N_0+N_2-N_1)\frac{n}{2}} (\prod_{a=1}^{N_0} x^{(0)}_a)^{n }
 (\prod_{a=1}^{N_2} x^{(2)}_a)^{-n}
 \end{align}
Here $\mathcal{J}=\{i_1,i_2, \cdots, i_n \}$ and the set $\mathcal{J}^c$ is the
complement of $\mathcal{J}$ in $\{1,\cdots, k\}$.
Then the sum over the residues with respect to $(x_{i_1},\cdots, x_{i_n})$ at
the poles of the first type \eqref{eq:poleszero} is given by
\begin{align}
&
\sum \underset{x_{i_n} }{\mathrm{Res}} \cdots \underset{x_{i_1} }{\mathrm{Res}} \, \frac{\mathcal{I}_k(x_1,\cdots, x_k)}{\prod_{\alpha=1}^k x_{\alpha}}
\nonumber \\
&=\frac{n! (k-n)!}{k!} (-1)^{(N_1-N_0)n}(\prod_{a=1}^{N_0} x^{(0)}_a)^{\frac{n}{2} }
 (\prod_{a=1}^{N_2} x^{(2)}_a)^{-\frac{n}{2}}
q^{ \frac{n}{4}(N_0+N_2-2N_1)}
\
\frac{   \mathcal{I}_{k-n}( \{x_i\}_{i \in \mathcal{J}^c})}{\prod_{\alpha \in \mathcal{J}^c} x_{\alpha}}
   \nonumber \\
& \times \frac{q^{\frac{n^2}{2}} }{n! (1-q)^n }
 \oint_{T^n_x} \prod_{\alpha=1}^n  \frac{d x_{i_{\alpha}}}{2 \pi {\rm i} x_{i_{\alpha}}}
\prod_{1 \le \alpha \neq \beta \le n } \frac{x_{i_{\alpha}}-x_{i_{\beta}} } {x_{i_{\alpha}}-q x_{i_{\beta}} }
\nonumber \\
&=\frac{n! (k-n)!}{k!} (-1)^{(N_1-N_0)n}
(\prod_{a=1}^{N_0} x^{(0)}_a)^{\frac{n}{2} }
 (\prod_{a=1}^{N_2} x^{(2)}_a)^{-\frac{n}{2}}
q^{ \frac{n}{4}(N_0+N_2-2N_1)}
\
 \frac{   \mathcal{I}_{k-n}( \{x_i\}_{i \in \mathcal{J}^c})}{\prod_{\alpha \in \mathcal{J}^c} x_{\alpha}}
  \frac{(-q^{\frac{1}{2}})^{n}}{(q;q)_{n}}
 \label{eq:interme1}
\end{align}
Here $(q;q)_n$ is a $q$-Pochhammer symbol defined by \eqref{eq:qfactorial}.
In the last line of \eqref{eq:interme1}, we used the following formula:
\begin{align}
&\frac{q^{\frac{n^2}{2}} }{n! (1-q)^n }
\sum \underset{x_{i_n} }{\mathrm{Res}} \cdots \underset{x_{i_1} }{\mathrm{Res}} \,
\prod_{\alpha =1}^n x^{-1}_{i_{\alpha}} \cdot \prod_{1 \le \alpha \neq \beta \le n } \frac{x_{i_{\alpha}}-x_{i_{\beta}} } {x_{i_{\alpha}}-q x_{i_{\beta}} }
\nonumber \\
&=
\frac{q^{\frac{n^2}{2}} }{n! (1-q)^n }
 \oint_{T^n_x} \prod_{\alpha=1}^n  \frac{d x_{i_{\alpha}}}{2 \pi {\rm i} x_{i_{\alpha}}}
\prod_{1 \le \alpha \neq \beta \le n } \frac{x_{i_{\alpha}}-x_{i_{\beta}} } {x_{i_{\alpha}}-q x_{i_{\beta}} }
=
 \frac{(-q^{\frac{1}{2}})^{n}}{(q;q)_{n}}\,.
\end{align}

The iterated residues of the second type for $\{ x_i \}_{i \in \mathcal{J}^c}$ in
\eqref{eq:interme1} give $Z^{(\zeta >0)}_{k-n}$.
By summing over $1 \le i_1 < \cdots < i_n \le k$ and $n=0,1,\cdots,k$, and taking
all possible residues at the first-type poles for $\{ x_i \}_{i \in \mathcal{J}}$
and at the second-type poles for $\{ x_i \}_{i \in \mathcal{J}^c}$, we obtain the
relation between $\mathcal{Z}_k$ and the $K$-theoretic vortex partition
functions:
\begin{align}
\mathcal{Z}_{k}
 =\left\{
\begin{array}{cl}
 \sum\limits_{n=0}^{k} (-1)^{(N_1-N_0)n}
\left(\prod_{a=1}^{N_0} x^{(0)}_a\right)^{\frac{n}{2}}
 \left(\prod_{a=1}^{N_2} x^{(2)}_a\right)^{-\frac{n}{2}}
q^{ \frac{n}{4}(N_0+N_2-2N_1)}
 \frac{(-q^{\frac{1}{2} })^{n}}{(q;q)_{n}}  {Z}^{(\zeta>0)}_{k-n}, & (\kappa=\frac{N_0-N_2}{2}  )\\
{Z}^{(\zeta>0)}_{k}, & (   \kappa  \neq \frac{N_0-N_2}{2} )
 \end{array}
\right. 
\label{eq:positive}
\end{align}

Next we consider evaluating the residues outside the torus $T^k_x$.
By changing the integration variables as $w_i=1/x_i$ for $i=1, \cdots, k$,
the contour integral is evaluated by the residues inside the torus $T^k_w$.

As in the case of the residues inside $T^k_x$,
the set of poles contributing to the residues for $w_i$ is classified into the
following two types.
The first type is given by
\begin{align}
&\{ w_{i_1}= q w_{i_2}, \, w_{i_2} = q w_{i_3},\, \cdots, w_{i_{n-1}}= q w_{i_n}, \,w_{i_n} =0 \}\,,\nonumber \\
&\{ w_{i_1}= q w_{i_3}, \, w_{i_2} = q^2 w_{i_3}, \,\cdots, w_{i_{n-1}}= q w_{i_n}, \,w_{i_n} = 0\}\,, \nonumber \\
&\quad \vdots \nonumber \\
&\{ w_{i_1}= 0, \, w_{i_2} = 0, \cdots, w_{i_{n-1}}= 0, \,w_{i_n} = 0\}\,.
\label{eq:poleszero2}
\end{align}
The poles of the second type are given by
\begin{align}
&\{ w_{i_{n+1}}= q w_{i_{n+2}}, \, w_{i_{n+2}} = q w_{i_{n+3}},\, \cdots, w_{i_{k-1}}= q w_{i_k}, \,w_{i_k} = (x^{(2)}_{j_1})^{-1} q^{\frac{1}{2}} \},\nonumber \\
&\{ w_{i_{n+1}}= q w_{i_{n+3}}, \, w_{i_{n+2}} = q^2 w_{i_{n+3}}, \,\cdots, w_{i_{k-1}}= q w_{i_k}, \,w_{i_k} = (x^{(2)}_{j_1})^{-1}  q^{\frac{1}{2}} \}, \nonumber \\
&\quad \vdots \nonumber \\
&\{ w_{i_{n+1}}= (x^{(2)}_{j_1})^{-1} q^{\frac{1}{2}}, \, w_{i_{n+2}} = (x^{(2)}_{j_2})^{-1} q^{\frac{1}{2}}, \cdots, w_{i_{k-1}}= (x^{(2)}_{j_{k-n-1}})^{-1} q^{\frac{1}{2}}, \,w_{i_{k}} = (x^{(2)}_{j_{k-n}})^{-1} q^{\frac{1}{2}} \}.
\label{eq:polesmass2}
\end{align}
Here $j_1, \cdots, j_{k-n} \in \{N_1+1,\cdots, N_2 \}$.
Note that, if $\kappa \neq \frac{N_2-N_0}{2}$, the residues at
\eqref{eq:poleszero2} are zero. In this case, the residues at
\eqref{eq:polesmass2} agree with $Z^{(\zeta <0)}_k$ under the identification of
the parameters \eqref{eq:idenpara}. On the other hand, if
$\kappa = \frac{N_2-N_0}{2}$, the residues at the first-type poles give
non-zero contributions. In a similar way as in \eqref{eq:interme1}, these
residues can be calculated as follows.
\begin{align}
&
\sum \underset{w_{i_n} }{\mathrm{Res}} \cdots \underset{w_{i_1} }{\mathrm{Res}}
\frac{\mathcal{I}_k(w^{-1}_1,\cdots, w^{-1}_k)}{\prod_{\alpha=1}^k w_{\alpha}}
\nonumber \\
&=\frac{n! (k-n)!}{k!} (-1)^{(N_2-N_1)n}(\prod_{a=1}^{N_0} x^{(0)}_a)^{-\frac{n}{2} }
 (\prod_{a=1}^{N_2} x^{(2)}_a)^{\frac{n}{2}}
q^{ \frac{n}{4}(2N_1-N_0-N_2)}
   \frac{   \mathcal{I}_{k-n}( \{ w^{-1}_{\alpha} \}_{\alpha \in \mathcal{J}^c} )}{\prod_{\alpha \in \mathcal{J}^c } w^{-1}_{\alpha}}
 \nonumber \\
& \times \frac{q^{\frac{n^2}{2}} }{n! (1-q)^n }   \oint_{T^n_w} \prod_{\alpha=1}^n  \frac{d w_{i_{\alpha}}}{2 \pi {\rm i} w_{i_{\alpha}}}  
\prod_{1 \le \alpha \neq \beta \le n } \frac{w_{i_{\alpha}}-w_{i_{\beta}} } {w_{i_{\alpha}}-q w_{i_{\beta}} }
\nonumber \\
&=\frac{n! (k-n)!}{k!} (-1)^{(N_2-N_1)n}(\prod_{a=1}^{N_0} x^{(0)}_a)^{-\frac{n}{2} }
 (\prod_{a=1}^{N_2} x^{(2)}_a)^{\frac{n}{2}}
q^{ \frac{n}{4}(2N_1-N_0-N_2)}
\
 \frac{   \mathcal{I}_{k-n}( \{ w^{-1}_{\alpha} \}_{\alpha \in \mathcal{J}^c} )}{\prod_{\alpha \in \mathcal{J}^c } w^{-1}_{\alpha}}
  \frac{(-q^{\frac{1}{2}})^{n}}{(q;q)_{n}}\,.
 \label{eq:interme2}
\end{align}
We evaluate the residues at \eqref{eq:polesmass2} and take the sum over all poles
of the first and second types. Then we obtain
\begin{align}
\mathcal{Z}_{k}
 =\left\{
\begin{array}{cl}
 \sum\limits_{n=0}^k (-1)^{(N_2-N_1)n}
(\prod_{a=1}^{N_0} x^{(0)}_a)^{-\frac{n}{2} }
 (\prod_{a=1}^{N_2} x^{(2)}_a)^{\frac{n}{2}}
q^{ \frac{n}{4}(2N_1-N_0-N_2)}
 \frac{(-q^{\frac{1}{2} })^{n}}{(q;q)_{n}}  {Z}^{(\zeta<0)}_{k-n}, & (\kappa=\frac{N_2-N_0}{2}  )\\
{Z}^{(\zeta<0)}_{k}, & (  \kappa \neq \frac{N_2-N_0}{2} )
 \end{array}
\right. 
\label{eq:negative}
\end{align}
The equality between \eqref{eq:positive} and \eqref{eq:negative} gives the
relation between $Z^{(\zeta>0)}_k$ and $Z^{(\zeta<0)}_k$, i.e., the
wall-crossing formula for the $K$-theoretic vortex partition functions. 

To express the wall-crossing formula in a compact form, we define the generating
function of the $k$-vortex partition functions by
\begin{align}
Z^{(\zeta)} =1+\sum_{k=1}^{\infty} z^{k} Z^{(\zeta)}_{k}\,.
\end{align}
For simplicity, we often refer to the generating function of the vortex partition
functions as the vortex partition function.
Then the wall-crossing formula is expressed as
\begin{align}
\sum_{k=0}^{\infty} z^k \mathcal{Z}_k&= \left(\sum_{n=0}^{\infty} \frac{z^n_+}{(q;q)_n} \right)^{\delta_{2 \kappa,N_0-N_2}} Z^{(\zeta>0)}
 \nonumber \\
&
=\left(\sum_{n=0}^{\infty} \frac{z^n_-}{(q;q)_n} \right)^{\delta_{2\kappa,N_2-N_0}} Z^{(\zeta<0)}\,,
\end{align}
where $z_{+}$ and $z_{-}$ are defined by
\begin{align} 
z_+
 &:=  
 (-1)^{N_1-N_0+1}(\prod_{a=1}^{N_0} x^{(0)}_a)^{\frac{1}{2} }
 (\prod_{a=1}^{N_2} x^{(2)}_a)^{-\frac{1}{2}} q^{ \frac{1}{4}(N_0+N_2-2N_1+2)}   z \,,\\
 z_-&:=(-1)^{N_1-N_2+1}(\prod_{a=1}^{N_0} x^{(0)}_a)^{-\frac{1}{2} }
 (\prod_{a=1}^{N_2} x^{(2)}_a)^{\frac{1}{2}}
q^{ -\frac{1}{4}(N_0+N_2-2N_1-2)}   z\,.
\end{align}
Using the $q$-binomial theorem \eqref{eq:qbinomi},
we obtain the following expression for the wall-crossing formula of the
$K$-theoretic vortex partition function in 3d $\mathcal{N}=2$ gauge theory with
a Chern-Simons term \cite{Hwang:2017kmk}:
\begin{align}
{(z_+;q)_{\infty}^{{\delta_{2 \kappa, N_0-N_2} }} } Z^{(\zeta>0)} ={(z_-;q)_{\infty}^{{\delta_{2 \kappa, N_2-N_0}}}} \, Z^{(\zeta<0)}.
\label{eq:WCformula1}
\end{align}

\subsection{Kajihara transformation as a wall-crossing formula in 3d $\mathcal{N}=2$ theory}
\label{sec:Kajihara}

We show that the wall-crossing formula \eqref{eq:WCformula1} for $N_2=N_0$ and
$\kappa=0$ agrees with the Kajihara transformation
\cite{kajihara2004euler}.
Using the identities \eqref{eq:id1inapp} and \eqref{eq:id2inapp}, the
$K$-theoretic vortex partition functions \eqref{eq:kvortexplus} and
\eqref{eq:kvortexminus}, together with \eqref{eq:idenpara}, can be expressed as
\begin{align}
Z^{(\zeta>0)}
 &= \sum_{k=0}^{\infty} 
  z_+^k 
 \sum_{k_1+\cdots+k_{N_1}=k} \,\,
  \prod_{1 \le  i < j \le N_1} \frac{x^{(2)}_i q^{ k_i}-x^{(2)}_j q^{ k_j} }{x^{(2)}_i-x^{(2)}_j}
 \nonumber \\
& \qquad \times \prod_{i, j=1}^{N_1}  \frac{(x^{(2)}_i / x^{(0)}_j ;q)_{k_i}}{ (q x^{(2)}_i / x^{(2)}_j;q )_{k_i}}  \cdot
\prod_{i =1}^{N_1} \prod_{j =N_1+1}^{N_2} \frac{( x^{(2)}_i / x^{(0)}_j ;q)_{k_i}}{(q x^{(2)}_i  /x^{(2)}_j;q )_{k_i}}\,, \\
Z^{(\zeta<0)}
 &= \sum_{k=0}^{\infty}
   z_{-}^{{k}} 
   \sum_{{k}_{N_1+1}+\cdots+{k}_{N_2}={k}}  \,\, \prod_{N_1+1 \le  i < j \le N_2}  \frac{(x^{(2)}_i)^{-1} q^{ {k}_i}-(x^{(2)}_j)^{-1} q^{ {k}_j} }{(x^{(2)}_i)^{-1}-(x^{(2)}_j)^{-1}}
  \nonumber \\
&\qquad \times 
 \prod_{i, j=N_1+1}^{N_2}  \frac{(q x^{(0)}_i / x^{(2)}_j ;q)_{{k}_i}}{ (q x^{(2)}_i / x^{(2)}_j;q )_{{k}_i}} \cdot
\prod_{i =N_1+1}^{N_2} \prod_{j =1}^{N_1} \frac{(q  x^{(0)}_j/ x^{(2)}_i  ;q)_{{k}_i}}{(q x^{(2)}_j  / x^{(2)}_i;q )_{{k}_i}} \,.
\end{align}

Next, we define ${\sf x}_i$, ${\sf y}_i$, ${\sf a}_i$, ${\sf b}_i$, ${\sf c}$,
and ${\sf z}$ by
\begin{align}
  x^{(2)}_i&=\left\{
\begin{array}{cl}
\frac{{\sf x}_i}{ {\sf x}_{N_1}}, & i \in \{1,\cdots, N_1 \}  \,,\\
 \frac{q \,{\sf y}_{N_2-N_1}}{c \, {\sf y}_{i-N_1}}, &   i \in \{N_1+1,\cdots,N_2 \} \,
 \end{array}
\right.
\label{eq:paraid1}
\\
  x^{(0)}_i&=\left\{
\begin{array}{cl}
\frac{{\sf x}_i}{{\sf x}_{N_1} {\sf a}_i} , & i \in \{1,\cdots, N_1 \}  \,,\\
  \frac{{\sf y}_{N_2-N_1}}{{\sf b}_{i-N_1}\,  {\sf y}_{i-N_1}}\,, &   i \in \{N_1+1,\cdots,N_2 \}\,,
 \end{array}
\right.
\label{eq:paraid2} \\
(-1)^{N_1-N_2+1} &(\prod_{a=1}^{N_2} x^{(0)}_a/x^{(2)}_a )^{\frac{1}{2} }
  q^{ \frac{1}{2}(N_2-N_1+1)} z =z_+={\sf z}\,.
\end{align}
Then the wall-crossing formula \eqref{eq:WCformula1} can be written as
\begin{align}
 &
 \sum_{k=0}^{\infty} {\sf z}^k  \sum_{k_1+\cdots+k_{N_1}=k}
 \prod_{1 \le i  < j \le N_1}
\frac{{\sf x}_i q^{ k_i }-{\sf x}_j q^{ k_j} }{{\sf x}_i-{\sf x}_j}  \cdot \prod_{i, j=1}^{N_1}  \frac{( {\sf a}_j {\sf x}_i / {\sf x}_j ;q)_{k_i}}{ (q\, {\sf x}_i / {\sf x}_j ;q )_{k_i}}
 \cdot \prod_{i =1}^{N_1} \prod_{j =1}^{N_2-N_1} \frac{({\sf b}_j  {\sf x}_i {\sf y}_j  / {\sf x}_{N_1}{\sf y}_{N_2-N_1} ;q)_{k_i}}{(c {\sf x}_i   {\sf y}_j/{\sf x}_{N_1} {\sf y}_{N_2-N_1};q )_{k_i}} \nonumber \\
&=\frac{({\sf z} \,c^{N_2-N_1} \prod_{i=1}^{N_1} {\sf a}_i \prod_{j=1}^{N_2-N_1} {\sf b}_j  ;q)_{\infty}}{({\sf z};q)_{\infty}}
\sum_{k=0}^{\infty} \Bigl( {\sf z}\, c^{N_2-N_1} \prod_{i=1}^{N_1} {\sf a}_i \prod_{j=1}^{N_2-N_1} {\sf b}_j  \Bigr)^{{k}} \sum_{{k}_1+\cdots+{k}_{N_2-N_1}={k}}
\nonumber \\
& \times
  \prod_{1 \le i < j \le N_2-N_1}
\frac{{\sf y}_i q^{ {k}_i}-{\sf y}_j q^{ k_j} }{{\sf y}_i-{\sf y}_j} \cdot \prod_{i, j=1}^{N_2-N_1} \frac{((c/ {\sf b}_j) {\sf y}_i / {\sf y}_j ;q)_{{k}_i}}{ (q\, {\sf y}_i / {\sf y}_j;q )_{{k}_i}}  \cdot
\prod_{i =1}^{N_2-N_1} \prod_{j =1}^{N_1} \frac{( (c /{\sf a}_j ){\sf y}_i {\sf x}_j  / {\sf y}_{N_2-N_1} {\sf x}_{N_1} ;q)_{{k}_i}}{(c {\sf y}_i   {\sf x}_j/{\sf y}_{N_2-N_1} {\sf x}_{N_1} ;q )_{{k}_i}} \nonumber \\
\end{align}
This agrees with the Kajihara transformation \cite{kajihara2004euler} for the
Kajihara-Noumi multiple hypergeometric series \cite{KajiharaNoumi2003}.
The wall-crossing formula for general $N_2$, $N_0$, and $\kappa$ can be regarded
as a generalization of the Kajihara transformation.

\section{Halln\"as-Langmann-Noumi-Rosengren formula as a wall-crossing formula}
\label{sec:HLNRformula}

\begin{figure}[thb]
  \centering
  \includegraphics[height=4.5cm]{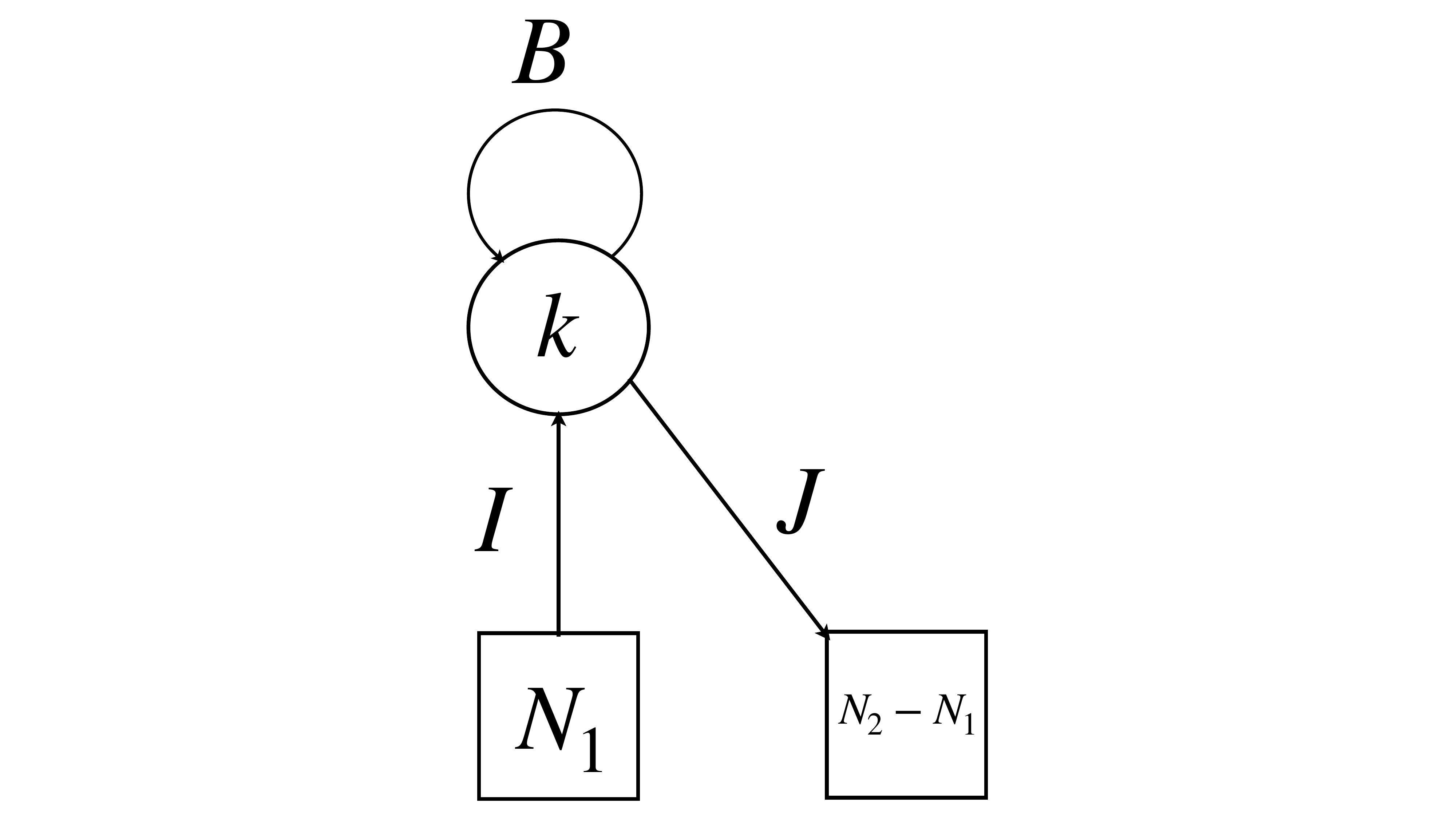}
  \caption{The quiver diagram representing the 1d $\mathcal{N}=4$ SQM associated
  with the $k$-vortex moduli for the 3d $\mathcal{N}=4$ gauge theory.
  The circle labeled k denotes the 1d $\mathcal{N}=4$ $U(k)$ vector multiplet.
  A solid arrow denotes a 1d $\mathcal{N}=4$ chiral multiplet. $B$, $I$, and
  $J$ denote the scalar fields in the 1d $\mathcal{N}=4$ chiral multiplets.
  The representations of the 1d multiplets are the same as in the 1d
  $\mathcal{N}=2$ case and are summarized in Table \ref{table:chargeN4}}
  \label{fig:handsaw2}
\end{figure}

\begin{table}[thb]
\begin{center}
\begin{tabular}{c || c c c  c c c}
			&	$U(k)$ &	$U(N_1)$ & $U(N_2-N_1)$ & $U(N_0)$ &  $U(1)_{\varepsilon}$
			 	\\
			 \hline
$ B$	 &	$\bm{adj}$	&	${\bm 1}$	& ${\bm 1}$ & ${\bm 1}$   & $-1$	\\
$ I$	 &	${\bm k}$	&	$\overline{\bm N_1}$	& ${\bm 1}$ & ${\bm 1}$ &  $-\frac{1}{2}$ \\
$ J$	 &	$\bar{\bm k}$	&	${\bm 1}$	& ${{\bm N_2}-{\bm N_1}}$ & ${\bm 1}$  & $-\frac{1}{2}$	\\
\end{tabular}
\caption{The representations of the scalars in the 1d $\mathcal{N}=4$ chiral
multiplets describing the $k$-vortex moduli space in the 3d $\mathcal{N}=4$
$U(N_1)$ gauge theory depicted by the quiver diagram in Figure
\ref{fig:handsaw2}. $B$, $I$, and $J$ are the scalar fields in the chiral
multiplets. $U(k)$ is the gauge group of the SQM. $U(N_1)$, $U(N_2-N_1)$,
$U(N_0)$, and $U(1)_{\varepsilon}$ are flavor symmetry groups.
}
\end{center}
\label{table:chargeN4}
\end{table}

In the previous section, we showed that the wall-crossing formula for the vortex
partition function in 3d $\mathcal{N}=2$ supersymmetric gauge theory agrees with
the Kajihara transformation.
In this section, we show that the wall-crossing formula for the vortex partition
function in 3d $\mathcal{N}=4$ $U(N_1)$ supersymmetric gauge theory agrees with
the trigonometric limit of the transformation formula of Halln\"as, Langmann,
Noumi, and Rosengren \cite{Halln_s_2022}.
The hypermultiplets in the $U(N_1)$ gauge theory have the same matter content as
the chiral multiplets in Table \ref{table:charge1} with $N_0=N_2$.
The 3d $\mathcal{N}=4$ $U(N_1)$ vector multiplet consists of a 3d $\mathcal{N}=2$
vector multiplet together with an additional chiral multiplet in the adjoint
representation of $U(N_1)$.

\subsection{Vortex partition function in 3d $\mathcal{N}=4$ gauge theory}
\label{sec:vortex3dN4}

First we briefly review the $K$-theoretic vortex partition function in 3d
$\mathcal{N}=4$ gauge theory.
As explained in Section \ref{section:Kvortex}, the partition function of the
$\mathcal{N}=2$ gauge theory on $S^1 \times \mathbb{R}^2$ reduces to the
$K$-theoretic vortex partition function given by the Witten index of 1d
$\mathcal{N}=2$ SQM.
Since the vortex equations are half-BPS equations, the partition function of the
3d $\mathcal{N}=4$ $U(N_1)$ gauge theory on $S^1 \times \mathbb{R}^2$ reduces to
the $K$-theoretic vortex partition function given by the Witten index of 1d
$\mathcal{N}=4$ SQM depicted by the handsaw quiver in Figure
\ref{fig:handsaw2}.

From the supersymmetric localization formula, the $K$-theoretic $k$-vortex
partition function in 3d $\mathcal{N}=4$ gauge theory is given by
\begin{align}
Z^{(\zeta)}_{k,(N_1,N_2)}&=
\frac{1}{k! (2 \sinh \frac{-m}{2})^k } \oint_{JK(\zeta)} \prod_{\alpha=1}^k \frac{d u_{\alpha}}{2 \pi {\rm i}} \left(
\prod_{1 \le \alpha \neq  \beta \le k }  \frac{2 \sinh \frac{u_{\alpha}-u_{\beta}}{2}  } {2 \sinh \frac{u_\alpha-u_{\beta}-m}{2} } \right)  
\left( \prod_{ \alpha, \beta=1 }^k  \frac{2 \sinh \frac{u_{\alpha}-u_{\beta}-m-\varepsilon}{2}  } {2 \sinh \frac{u_\alpha-u_{\beta}-\varepsilon}{2} } \right)
\nonumber \\
& \qquad  \times 
\prod_{\alpha=1}^{k} \prod_{j=1}^{N_1}   \frac{2\sinh \left( \frac{u_\alpha-m_j-m -\varepsilon/2 }{2} \right) }{ 2 \sinh \left( \frac{u_\alpha-m_j -\varepsilon/2 }{2} \right)}
\prod_{j =N_1+1}^{N_2}  \frac{2 \sinh \left( \frac{-u_\alpha+m_j-m-\varepsilon/2}{2} \right)}{ 2 \sinh \left( \frac{-u_\alpha+m_j-\varepsilon/2}{2} \right)} \,.
 \label{eq:contour1N4}
\end{align}
Here again, the parameters $\{ m_i \}_{i =1}^{N_1}$, $\{ m_i \}_{i=N_1+1}^{N_2}$,
and $\varepsilon$ are the chemical potentials for the 1d flavor symmetry groups
$U(N_1)$, $U(N_2-N_1)$, and $U(1)_{\varepsilon}$, respectively.
The parameter $m$ is a chemical potential for a linear combination of the
generator of the $U(1)$ part of the $SU(2)$ R-symmetry and the generator of the
$U(1)$ rotation of the time circle in 1d $\mathcal{N}=4$ SQM.

Since the residues coming from the pole $u_{\alpha}-u_{\beta}-m=0$ are zero, in
the positive FI-parameter region, the poles contributing to the JK residue of
\eqref{eq:contour1N4} are the same as those in \eqref{eq:hyperplanes1}.
Similarly, the JK residue of \eqref{eq:contour1N4} in the negative FI-parameter
region is evaluated at \eqref{eq:hyperplanes2}. Then the $k$-vortex partition
function for 3d $\mathcal{N}=4$ gauge theory is given by
{\small
\begin{align}
 Z^{(\zeta>0)}_{k,(N_1,N_2)}&=\sum_{k_1+\cdots +k_{N_1}=k} \prod_{i,j =1}^{N_1}  \prod_{\ell_i=1}^{k_i}
  \frac{ 2 \sinh \left( \frac{m_i-m_j -m+(\ell_i -k_j-1)\varepsilon }{2} \right) } { 2 \sinh \left( \frac{m_i-m_j +(\ell_i -k_j-1)\varepsilon }{2} \right) } \cdot
 \prod_{i =1}^{N_1} \prod_{j =N_1+1}^{N_2} \prod_{\ell_i=1}^{k_i}  \frac{2 \sinh \left( \frac{-m_i+m_j-m-\ell_i\varepsilon}{2} \right)}
 { 2 \sinh \left( \frac{-m_i+m_j-\ell_i\varepsilon}{2} \right)}   
 \label{eq:kvortexplusN4}, \\
 Z^{(\zeta<0)}_{k, (N_1,N_2)}&=\sum_{k_{N_1+1}+\cdots +k_{N_2}=k} \prod_{i,j =N_1+1}^{N_2}  \prod_{\ell_i=1}^{k_i} \frac{ 2 \sinh \left( \frac{-m_i+m_j -m+(\ell_i -k_j-1)\varepsilon }{2} \right) } { 2 \sinh \left( \frac{-m_i+m_j +(\ell_j -k_i-1)\varepsilon }{2} \right) } \cdot
 \prod_{i =N_1+1}^{N_2} \prod_{j =1}^{N_1} \prod_{\ell_i=1}^{k_i}  \frac{2 \sinh \left( \frac{m_i-m_j-m-\ell_i \varepsilon}{2} \right)}
 { 2 \sinh \left( \frac{m_i-m_j-\ell_i\varepsilon}{2} \right)}.
 \label{eq:kvortexminusN4}
 \end{align}
}

\subsection{Halln\"as-Langmann-Noumi-Rosengren formula as a wall-crossing formula in 3d $\mathcal{N}=4$ theory}
\label{sec:WCHLNR}

In Section \ref{sec:wallcross}, we derived the wall-crossing formula in terms of
the contour integral \eqref{eq:integrand2}.
First we define the generating function of the $K$-theoretic $k$-vortex
partition functions for 3d $\mathcal{N}=4$ gauge theory by
\begin{align}
Z^{(\zeta)}_{(N_1,N_2)}(x, \tilde{x}, t,q,z) = 1+\sum_{k=1}^{\infty} z^{k} Z^{(\zeta)}_{k,(N_1,N_2)}(x, \tilde{x}, t,q)\,,
\label{eq:generatN4}
\end{align}
where $x_i$, $\tilde{x}_i$, and $t$ are defined by
\begin{align}
x_i=e^{m_i}, \,\, \tilde{x}_j=e^{m_{j}}, \,\, t=e^{m}  \,,
\end{align}
for $i \in \{1,\cdots, N_1 \}$, $j \in \{N_1+1,\cdots, N_2 \}$.
$q$ is again the exponentiated $\Omega$-background parameter $q=e^{\varepsilon}$.

In this subsection, we review the large-mass limit applied in
\cite{Hwang:2015wna} to derive the wall-crossing formula for the 3d
$\mathcal{N}=4$ gauge theory. In order to prove the wall-crossing formula, we consider the case $N_2=2N_1$.
From the expressions \eqref{eq:kvortexplusN4} and \eqref{eq:kvortexminusN4}, the
following relation holds between the $K$-theoretic vortex partition functions
under the exchange of the variables $m_i \leftrightarrow -m_{i+N_1}$ for
$i=1,\cdots, N_1$:
\begin{align}
 Z^{(\zeta>0)}_{(N_1,2 N_1)}(x, \tilde{x} ,t,q,z)= Z^{(\zeta>0)}_{(N_1,2 N_1)}(\tilde{x}^{-1},x^{-1} ,t,q,z)
 \label{eq:2N1eqN2}
\end{align}
We take a large negative mass limit, ${\rm Re}(m_{2N_1}) \to -\infty$, in
\eqref{eq:2N1eqN2}.
The limit $\tilde{x}_{ N_1 } \to 0$ on the left-hand side of \eqref{eq:2N1eqN2}
is given by
\begin{align}
\lim_{\tilde{x}_{ N_1 } \to 0} Z^{(\zeta>0)}_{(N_1,2 N_1)}(x, \tilde{x} ,t,q, z)= Z^{(\zeta>0)}_{(N_1,2 N_1-1)}(x, \tilde{x} ,t,q, z t^{\frac{1}{2}})
\label{eq:2N1m1qN2}
\end{align}
On the other hand, the limit $\tilde{x}_{N_1} \to 0$ on the right-hand side of
\eqref{eq:2N1eqN2} is given by
\begin{align}
\lim_{\tilde{x}_{ N_1 } \to 0} Z^{(\zeta>0)}_{(N_1,2 N_1)}(\tilde{x}^{-1},{x}^{-1}  ,t,q, z )&=
Z^{(\zeta)}_{(1,1)}(t,q, z t^{\frac{1}{2}}) Z^{(\zeta>0)}_{(N_1-1,2 N_1-1)}(\tilde{x}^{-1},{x}^{-1}  ,t,q, z t^{\frac{1}{2}} )
 \nonumber \\
&=
\frac{( z q t ;q)_{\infty}}{(z  ;q)_{\infty}}  Z^{(\zeta>0)}_{(N_1-1,2 N_1-1)}(\tilde{x}^{-1},{x}^{-1}  ,t,q, z t^{\frac{1}{2}} )
\end{align}
Here, we used the following identity:
\begin{align}
Z^{(\zeta)}_{(1,1)}(t,q, z ) &=\sum_{k=0}^{\infty} z^k \prod_{\ell=1}^{k}
 \frac{ 2 \sinh \left( \frac{-m+(\ell-k-1)\varepsilon }{2} \right) } { 2 \sinh \left( \frac{(\ell -k-1)\varepsilon }{2} \right) }
 =\sum_{k=0}^{\infty} (t^{-\frac{1}{2}}z)^k \frac{(t q;q)_k}{(q;q)_k}  \nonumber \\
 &=\frac{( z q t^{\frac{1}{2}};q)_{\infty}}{(z t^{-\frac{1}{2}};q)_{\infty}} \,.
\end{align}
Note that the vortex partition function $Z^{(\zeta)}_{(1,1)}(t,q, z )$ is
independent of the sign of the 1d FI parameter $\zeta$.
In a similar way, by repeating this procedure, namely taking the limit
$\tilde{x}_s \to 0$ for $s=2 N_1, 2 N_1-1, \cdots, 2N_1-N_2$ in
\eqref{eq:2N1eqN2}, we obtain the following relation:
\begin{align}
   Z^{(\zeta>0)}_{(N_1,N_2)}(x,\tilde{x},t,q,z t^{\frac{2N_1-N_2}{2}}) &=\prod_{s=1}^{2N_1-N_2}\frac{(z q \, t^{s} ;q)_{\infty}}{(z \, t^{s-1} ;q)_{\infty}} \, \cdot Z^{(\zeta>0)}_{(N_2-N_1, N_2)}(\tilde{x}^{-1}, x^{-1}, t,q,z t^{\frac{2N_1-N_2}{2}})\, .
  \label{eq:wallcrossingN40}
\end{align}
Therefore, we obtain the wall-crossing formula for 3d $\mathcal{N}=4$ $U(N_1)$
gauge theory with $N_2$ hypermultiplets \cite{Hwang:2015wna}
\begin{align}
  Z^{(\zeta>0)}_{(N_1,N_2)}(x,\tilde{x},t,q,z) & =\prod_{s=1}^{2N_1-N_2}\frac{(z q \, t^{s+\frac{N_2-2N_1}{2}} ;q)_{\infty}}{(z \, t^{s-1+\frac{N_2-2N_1}{2}} ;q)_{\infty}} \, \cdot Z^{(\zeta<0)}_{(N_1, N_2)}(x, \tilde{x}, t,q,z),
  \label{eq:wallcrossingN4}
\end{align}
where we redefine $z \mapsto z t^{N_2-2N_1}$ and use the relation
\begin{align}
  Z^{(\zeta>0)}_{(N_2- N_1, N_2)}(\tilde{x}^{-1},{x}^{-1} ,t,q)=Z^{(\zeta<0)}_{(N_1, N_2)}(x, \tilde{x} ,t,q)\,.
\end{align}

In \eqref{eq:wallcrossingN4}, we set
\begin{align}
  x_i&=e^{m_i}={\sf x}_i, \,\, i  \in \{1,\cdots, N_1 \}\,,
\label{eq:paraidN41}
\\
  \tilde{x}^{-1}_{i-N_1} t q&=e^{-m_i+m-\varepsilon}={\sf y}_{i-N_1}, \,\,  i \in \{N_1+1,\cdots, N_2 \}\,,
\label{eq:paraidN42} \\
t&= {\sf t}/{q}\,,\\
z t^{\frac{2N_1-N_2}{2}}&= z e^{\frac{2N_1-N_2}{2} m}={\sf z}\,.
\end{align}
Then the wall-crossing formula \eqref{eq:wallcrossingN4} is expressed as
\begin{align}
&\sum_{k=0}^{\infty} {\sf z}^k \sum_{k_1+\cdots+k_{N_1}=k}\prod_{i,j=1}^{N_1} \frac{(q^{1-k_j} {\sf x}_i/({\sf t} {\sf x}_j);q )_{k_i}}{(q^{-k_j} {\sf x}_i/{\sf x}_j;q )_{k_i}}
\prod_{i=1}^{N_1} \prod_{j=1}^{N_2-N_1} \frac{( {\sf x}_i  {\sf y}_j; q )_{k_i}}{(q {\sf x}_i{\sf y}_j /{\sf t};q )_{k_i}} \nonumber \\
&=\prod_{s=1}^{2N_1-N_2}\frac{({\sf z} q^{s}/{\sf t}^{s-1};q )_{\infty}}{({\sf z} q^s/{\sf t}^{s};q )_{\infty}}\sum_{k=0}^{\infty} \left( \frac{q^{2N_1-N_2} }{{\sf t}^{2N_1-N_2}}  {\sf z} \right)^k
\nonumber \\
& \qquad \times \sum_{k_1+\cdots+k_{N_2-N_1}=k}\prod_{i,j=1}^{N_2-N_1} \frac{(q^{1-k_j} {\sf y}_i/({\sf t} {\sf y}_j);q )_{k_i}}{(q^{-k_j} {\sf y}_i/{\sf y}_j;q )_{k_i}}
\prod_{i=1}^{N_2-N_1} \prod_{j=1}^{N_1} \frac{( {\sf y}_i  {\sf x}_j; q )_{k_i}}{(q {\sf y}_i{\sf x}_j /{\sf t};q )_{k_i}}\,.
\end{align}
This agrees with the trigonometric limit of the formula in
\cite{Halln_s_2022}. For example, see Eq.~(14) of
\cite{Ohkawa:2022ebc}.

\subsection{$K$-theoretic vortex partition function as equivariant $\chi_t$-genus}
\label{sec:chitgenus}

Since the Fermi multiplet is absent in the 1d $\mathcal{N}=4$ SQM, and a 1d
$\mathcal{N}=4$ chiral multiplet contains additional fermions compared with a 1d
$\mathcal{N}=2$ chiral multiplet, the numerator of the integrand in
\eqref{eq:contour1N4} has changed from that in the $\mathcal{N}=2$ case
\eqref{eq:contour1}. The fermions determine the characteristic class that
defines an index of the handsaw quiver variety. In particular, it is known that
the Witten index of the 1d $\mathcal{N}=4$ SQM gives the equivariant
$\chi_t$-genus of the moduli space of Higgs branch vacua of the SQM.

Let us explicitly look at the agreement between the $K$-theoretic vortex
partition function and the $\chi_t$-genus of the handsaw quiver variety.
We evaluate the equivariant character of the handsaw quiver variety of type
$A_1$ \cite{Yoshida:2011au, Bonelli:2011fq, Fujitsuka:2013fga}.
Elements of the complexification of the maximal tori of $U(k)$ and
$U(N_1) \times U(N_2-N_1) \times U(1)_{\varepsilon}$ act on
\eqref{eq:Higgsvac1GL} as
\begin{align}
&g: (B, I,J) \mapsto (g B g^{-1}, g I, J g^{-1}),\nonumber \\
 & g_1:(B,I,J) \mapsto (B,I g_1,J),  \nonumber \\
& g_2:(B,I,J) \mapsto (B,I , g_2 J), 
\\
& g_3: (B,I,J) \mapsto ( g_3 B , I ,  J ),  \nonumber
\end{align}
where
\begin{align}
 g&=\mathrm{diag} (e^{u_1},\cdots, e^{u_{k}}) \in (\mathbb{C}^{*})^{k}\,, \\
 g_1&=\mathrm{diag} (e^{{m_1}+\varepsilon/2},\cdots, e^{m_{N_1}+\varepsilon/2}) \in (\mathbb{C}^{*})^{N_1}\,, \\
 g_2&=\mathrm{diag} (e^{m_{N_{1}}-\varepsilon/2},\cdots, e^{m_{N_2}-\varepsilon/2}) \in (\mathbb{C}^{*})^{N_2-N_1}\,, \\
 g_3&=e^{-\varepsilon} \in \mathbb{C}^{*}\,.
 \end{align}
Then the fixed-point condition of the infinitesimal transformation generating
$g_i$, $i=1,2,3$, up to a $(\mathbb{C}^{*})^k$ transformation is given by
\begin{align}
\left(u_{\alpha}-u_{\beta} -\varepsilon \right)B_{\alpha, \beta}=0,  \,\,
\left(u_{\alpha} -m_i-\frac{\varepsilon}{2}  \right) I_{\alpha, i}=0, \,\,
\left(-u_{\alpha} +m_j-\frac{\varepsilon}{2}  \right) J_{j, \alpha}=0\,,
\label{eq:fixedcond}
\end{align}
with $\alpha, \beta  \in \{ 1,\cdots, k \}$, $i \in \{1,\cdots, N_1 \}$, and
$j\in \{N_1+1,\cdots, N_2 \}$.
Thus, the fixed-point condition is the same as either \eqref{eq:hyperplanes1} or
\eqref{eq:hyperplanes2}. First, we consider the case of
\eqref{eq:hyperplanes1}, in which the $I_{\alpha, i}$ have non-zero values,
which corresponds to $\zeta >0$.
In this case, the fixed-point condition \eqref{eq:fixedcond} is solved by
$u_{\alpha}=m_i+(\ell_i-1/2) \varepsilon$ with
$\ell_i \in \{ 1,\cdots, k_i \}$ and $i \in \{ 1,\cdots, N_1 \}$.
Each fiber of the tangent bundle of the handsaw quiver variety
$T\mathcal{M}^{(\zeta >0)}_{k, N_1,N_2}$ is given by
\begin{align}
\mathrm{End}(V) \otimes Q \oplus \mathrm{Hom} (W^{(1)}, V) \oplus \mathrm{Hom} (V, W^{(2)})/ \mathrm{End}(V)\,,
\end{align}
where $Q$ is a one-dimensional vector space on which $e^{-\varepsilon}$ acts.
Then the equivariant character $Ch(T\mathcal{M}^{(\zeta >0)}_{k, N_1,N_2})$ is
computed as
\begin{align}
{Ch}(T\mathcal{M}^{(\zeta >0)}_{k, N_1,N_2})
&= V \times V^* \times (Q-1) + V \times (W^{(1)})^* +W^{(2)}  \times V^*  \nonumber \\
&=\sum_{i,j=1}^{N_1} \sum_{\ell_i=1}^{k_i} e^{m_i -m_j+(\ell_i-k_j-1)\varepsilon}+
\sum_{i=1}^{N_1} \sum_{j=N_1+1}^{N_2}  \sum_{\ell_i=1}^{k_i} e^{-m_i +m_j-\ell_i\varepsilon}
\label{eq:character0}
\end{align}
Here, allowing for a slight abuse of notation, we denote each vector space and
its character by the same symbol. The characters are given by
\begin{align}
 V =\sum_{i=1}^{N_1} \sum_{\ell_i=1}^{k_i} e^{m_i+(\ell_i-1/2) \varepsilon},  \,\,W^{(1)} =\sum_{i=1}^{N_1}e^{{m_i}+\varepsilon/2}, \,\,W^{(2)} =\sum_{j=N_1+1}^{N_2}e^{m_j-\varepsilon/2}, \,\, Q=e^{-\varepsilon}.
 \label{eq:fixedcond0}
\end{align}
Note that the fixed points are classified by the partitions of the integer $k$
into $N_1$ non-negative integers $k_i$, for $i=1,\cdots, N_1$.

Now we relate the $K$-theoretic vortex partition function obtained by the
supersymmetric localization of the Witten index to the equivariant
$\chi_t$-genus:
\begin{align}
\chi_t (M)&= \int_M \mathrm{ch}(\wedge_{-t} TM^*) {\rm td}(M) \nonumber \\
&=\int_M \prod_{i} (1-t e^{-{\tt x}_i}) \frac{{\tt x}_i}{ 1-e^{-{\tt x}_i}}\,.
\end{align}
Here ${\tt x}_i$ are the Chern roots of the tangent bundle $TM$.
${\rm ch}(E)$ and ${\rm td}(M)$ are the Chern character of a vector bundle $E$
and the Todd class of $M$, respectively.
$\wedge_{t} E$ is the generating series of exterior powers of a vector bundle
$E$ (in $K$-theory) given by
\begin{align}
\wedge_{t} E= 1+ t  E + t^2 \wedge^2 E+ \cdots\,.
\end{align}
From the fixed-point formula, see for example \cite{Hollowood:2003cv}, the
$\chi_t$-genus is written as
\begin{align}
\chi_t (M)
&=\sum_{p:{\rm fixed}} \prod_{i} \frac{(1-t e^{-w_{i,p}}) \frac{w_{i,p} }{1-e^{-w_{i,p}} }}{\prod_{i }w_{i,p}} \nonumber  \\
&=\sum_{p:{\rm fixed}} \prod_{i} \frac{1-t e^{-w_{i,p}}}{ 1-e^{-w_{i,p}}}\,. 
\label{eq:fixedptformula}
\end{align}
Here $Ch(TM)=\sum_i e^{w_{i,p}}$ denotes the equivariant character at a fixed
point $p$. The sum $\sum_{p:{\rm fixed}}$ is taken over the fixed points $p$ of
the torus action.
From the equivariant character \eqref{eq:character0}, the $\chi_t$-genus of the
handsaw quiver variety is written as
\begin{align}
\chi_t (\mathcal{M}^{(\zeta>0)}_{k,N_1,N_2})
&=\sum_{k_1+\cdots +k_{N_1}=k} \prod_{i,j =1}^{N_1}  \prod_{\ell_i=1}^{k_i}
  \frac{ 1-  t e^{-m_i+m_j -(\ell_i -k_j-1)\varepsilon }  } { 1-   e^{-m_i+m_j -(\ell_i -k_j-1)\varepsilon }} \cdot
 \prod_{i =1}^{N_1} \prod_{j =N_1+1}^{N_2} \prod_{\ell_i=1}^{k_i}  \frac{1- t  e^{m_i-m_j+\ell_i\varepsilon}}
 { 1-  e^{m_i-m_j+\ell_i\varepsilon}}  \nonumber   \\
 &= t^{-\frac{N_2 k}{2}} Z^{(\zeta>0)}_{k, (N_1,N_2)} , \quad (t=e^{m}).
 \end{align}
Therefore the equivariant $\chi_t$-genus of
$\mathcal{M}^{(\zeta >0)}_{k,N_1,N_2}$ agrees with the $K$-theoretic
$k$-vortex partition function with positive FI parameter up to a factor
$t^{-\frac{N_2 k}{2}}$.

Next we consider the case where the fixed-point condition is given by
\eqref{eq:hyperplanes2}, which is solved by \eqref{eq:solu2}.
In this case, $J_{j,\alpha}$ have non-zero components, which means that the FI
parameter is negative: $\zeta <0$. Then the equivariant character of the
tangent space is computed as
\begin{align}
{Ch}(T\mathcal{M}^{(\zeta <0)}_{k, N_1,N_2})
&= V \times V^* \times (Q-1) + V \times (W^{(1)})^* +W^{(2)}  \times V^*  \nonumber \\
&=\sum_{i,j=N_1+1}^{N_2} \sum_{\ell_i=1}^{k_i} e^{-m_i +m_j+(\ell_i-k_j-1)\varepsilon}+
\sum_{i=N_1+1}^{N_2} \sum_{j=1}^{N_1}   \sum_{\ell_i=1}^{k_i} e^{m_i -m_j-\ell_i\varepsilon}\,,
\label{eq:character01}
\end{align}
where the characters of the spaces $V$ and $W^{(i)}$ are given by
\begin{align}
 V =\sum_{i=N_1+1}^{N_2} \sum_{\ell_i=1}^{k_i} e^{m_i-(\ell_i-1/2) \varepsilon},  \,\,W^{(1)} =\sum_{i=1}^{N_1}e^{{m_i}+\varepsilon/2}, \,\,W^{(2)} =\sum_{j=N_1+1}^{N_2}e^{m_j-\varepsilon/2}, \,\, Q=e^{-\varepsilon}\,.
 \label{eq:fixedcond01}
\end{align}
From the fixed-point formula \eqref{eq:fixedptformula}, we again find agreement:
\begin{align}
\chi_t (\mathcal{M}^{(\zeta<0)}_{k,N_1,N_2})
&=\sum_{k_{N_1+1}+\cdots +k_{N_2}=k} \prod_{i,j =N_1+1}^{N_2}  \prod_{\ell_i=1}^{k_i}
  \frac{ 1-  t e^{m_i-m_j -(\ell_i -k_j-1)\varepsilon }  } { 1-   e^{m_i-m_j -(\ell_i -k_j-1)\varepsilon }} \cdot
 \prod_{i =N_1+1}^{N_2} \prod_{j =1}^{N_1} \prod_{\ell_i=1}^{k_i}  \frac{1- t  e^{-m_i+m_j+\ell_i\varepsilon}}
 { 1-  e^{-m_i+m_j+\ell_i\varepsilon}}  \nonumber   \\
 &= t^{-\frac{N_2 k}{2}} Z^{(\zeta<0)}_{k, (N_1,N_2)} , \quad (t=e^{m}).
 \label{eq:chivor2}
 \end{align}
The difference between the $\chi_t$-genera and the vortex partition functions
can be absorbed into the redefinition $z \to z t^{\frac{N_2 }{2}}$ in
\eqref{eq:generatN4}, or eliminated by turning on a background flavor 1d
Chern-Simons term: $t^{\frac{N_2 k}{2}}$. Therefore, we have shown that the
wall-crossing formula of the $K$-theoretic vortex partition function for 3d
$\mathcal{N}=4$ gauge theory is the same as that of the $\chi_t$-genus of the
handsaw quiver variety of type $A_1$.

 Next, we consider the vortex partition functions for 3d $\mathcal{N}=2$ theory.
Since the computation is parallel to that in the above 3d $\mathcal{N}=4$ case,
we briefly comment on the geometric interpretation of the $K$-theoretic vortex
partition function for 3d $\mathcal{N}=2$ gauge theory.
We consider the vector bundle ${\cal E}$, whose fiber at each point is given by
$V$ modulo $GL(k,\mathbb{C})$, and consider the Chern character
$\mathrm{ch} (\otimes_{i=1}^{N_0} \wedge_{(x^{(0)}_i q)^{-1}} {\cal E} )$.
Then the index is written as
\begin{align}
&\int_{\mathcal{M}^{(\zeta)}_{k,N_1,N_2}} e^{\kappa^{\prime} c_1} \mathrm{ch} ( \otimes_{i=1}^{N_0} \wedge_{-(x^{(0)}_i q^{1/2})^{-1}} {\cal E} ) {\rm td}(\mathcal{M}^{(\zeta)}_{k,N_1,N_2}) \nonumber \\
&\qquad \qquad =
q^{\frac{k^2}{2}} \Bigl[q^{\frac{N_0+N_2}{2}} \prod_{i=1}^{N_0} x^{(0)}_i \cdot  \prod_{i=1}^{N_1 } x^{(2)}_i \cdot \prod_{i=N_1+1}^{N_2 }(x^{(2)}_i)^{-1})\Bigr]^{\frac{k}{2}} \,Z^{(\zeta)}_{k}  \,
\label{eq:indth}
\end{align}
Here $\kappa^{\prime}=\kappa+\frac{N_2-N_0-2N_1}{2}$ and $c_1$ is the first
Chern class.
Again, the prefactor given by the $k/2$-th power of $x^{(0)}_i$, $x_i^{(2)}$,
and $q$ is absorbed into the redefinition of $z$ in the generating function.
On the other hand, $q^{\frac{k^2}{2}}$ cannot be absorbed into any redefinition
of parameters.
Thus, the wall-crossing formula of the index \eqref{eq:indth} is derived from the
wall-crossing formula of the $k$-vortex partition function, but is not exactly
the same as the generating function: rather, it differs from it by a factor of
$q^{k^{2}/2}$.

\subsection{Elliptic analogue of wall-crossing formula}
\label{sec:elliptic}

In this section, we briefly comment on an elliptic analogue of the
wall-crossing formula.
First we consider the 4d $\mathcal{N}=2$ gauge theory on
$T^2 \times \mathbb{R}^2$, whose dimensional reduction gives the 3d
$\mathcal{N}=4$ gauge theory considered in Section \ref{sec:HLNRformula}.
The $k$-vortex partition function for the 4d $\mathcal{N}=2$ gauge theory is
given by the elliptic genus of the 2d $\mathcal{N}=(2,2)$ gauge theory on
$T^2$, where the matter content is given by the 2d lift of the 1d
$\mathcal{N}=4$ SQM in Figure \ref{fig:handsaw2}. The elliptic genus is again
evaluated using the supersymmetric localization formula in
\cite{Benini:2013nda, Benini:2013xpa}, and is written as
{\small
\begin{align}
 Z^{(\zeta>0)}_{4d, k,(N_1,N_2)}&=\sum_{k_1+\cdots +k_{N_1}=k} \prod_{i,j =1}^{N_1}  \prod_{\ell_i=1}^{k_i}
  \frac{ \theta_1(  m_{ij} -m+(\ell_i -k_j-1)\varepsilon ) } { \theta_1( m_{ij} +(\ell_i -k_j-1)\varepsilon )} \cdot
 \prod_{i =1}^{N_1} \prod_{j =N_1+1}^{N_2} \prod_{\ell_i=1}^{k_i}  \frac{\theta_1( m_{ji}-m-\ell_i\varepsilon )}
 { \theta_1( m_{j,i}-\ell_i\varepsilon )}   
 \label{eq:kvortexplus4dN4} \\
 Z^{(\zeta<0)}_{4d, k, (N_1,N_2)}&=\sum_{k_{N_1+1}+\cdots +k_{N_2}=k} \prod_{i,j =N_1+1}^{N_2}  \prod_{\ell_i=1}^{k_i} \frac{ \theta_1( m_{ji} -m+(\ell_i -k_j-1)\varepsilon ) } { \theta_1( m_{ji} +(\ell_j -k_i-1)\varepsilon ) } \cdot
 \prod_{i =N_1+1}^{N_2} \prod_{j =1}^{N_1} \prod_{\ell_i=1}^{k_i}  \frac{\theta_1(m_{ij}-m-\ell_i \varepsilon)}
 { \theta_1( m_{ij}-\ell_i\varepsilon )} 
 \label{eq:kvortexminus4dN4}
 \end{align}
}
Here $m_{ij}=m_i-m_j$, and $\zeta$ is a reference vector which plays the same
role as the FI parameter in the JK residue computation.
$\theta_1(z)$ is a Jacobi theta function defined by
\begin{align}
\theta_1(z)=\theta_1(\tau | z)=-{\rm i} y^{\frac{1}{2}} \prod_{k=1}^{\infty} (1-{\sf q}^k)(1-y {\sf q}^k) (1-y^{-1} {\sf q}^{k-1})\,,
\end{align}
where ${\sf q}=e^{2\pi {\rm i} \tau}$ and $y=e^{2 \pi {\rm i} z}$. $\tau$ is the
modulus of the two-dimensional torus $T^2$.
As explained in \cite{Benini:2013nda, Benini:2013xpa}, the elliptic genus does
not exhibit non-trivial wall-crossing phenomena. Thus the elliptic analogue of
$K$-theoretic vortex partition functions evaluated in two regions coincide with
each other up to a sign:
\begin{align}
 Z^{(\zeta>0)}_{4d, k,(N_1,N_2)}&=
 Z^{(\zeta<0)}_{4d, k, (N_1,N_2)}\,.
 \end{align}

Next we give a remark on the elliptic lift of the 3d $\mathcal{N}=2$ theory
without a Chern-Simons term.
In this case, the elliptic lift of the 1d $\mathcal{N}=2$ SQM for the
$k$-vortex moduli is given by a 2d $\mathcal{N}=(0,2)$ gauge theory on $T^2$,
which has a gauge anomaly and is inconsistent as a quantum field theory.
It would be interesting to find the elliptic lift of the Kajihara
transformation realized as an elliptic genus.

\section{Two-dimensional/cohomological/rational limit}
\label{sec:2dlimit}

In this section we take the zero-radius limit of the time circle $S^1$ and show
that the wall-crossing formulas reduce to the wall-crossing formulas for the
vortex partition functions in 2d $\mathcal{N}=(2,2)$ and
$\mathcal{N}=(2,2)^*$ gauge theories on $\mathbb{R}^2$.
Since an exponentiated chemical potential $e^{m_i}$ is a BPS background Wilson
loop for a $U(1)$ flavor symmetry, in the 2d limit, we may rescale it as
$m_i \to -\beta m_i$ and take the limit $\beta \to 0$ while keeping the new
$m_i$ finite. Here $\beta$ is the circumference of $S^1$, and the new $m_i$ is
a twisted mass in the 2d theory.
Let us rescale all the chemical potentials and take the zero-radius limit.

First we consider the wall-crossing formula for the 2d
$\mathcal{N}=(2,2)$ gauge theory obtained from the 2d limit of the 3d
$\mathcal{N}=2$ gauge theory with $N_2=N_0$ and $\kappa=0$.
In the above limit, $\beta \to 0$, the $K$-theoretic vortex partition functions
reduce to
{\small
 \begin{align}
&Z^{(\zeta>0)}_{2d, k}:=\lim_{\beta \to 0}Z^{(\zeta>0)}_{k} \nonumber \\
&=\sum_{k_1+\cdots +k_{N_1}=k} \frac{ (-1)^{N_1k} \prod_{i =1}^{N_1} \prod_{j=1}^{N_{2}}   \prod_{\ell_i=1}^{k_i}   \left(m_i-\tilde{m}_j+ (\ell_i-1) \varepsilon \right)}
 {\prod_{i,j =1}^{N_1}  \prod_{\ell_i=1}^{k_i}   \left( m_j-m_i -(\ell_i -k_j-1)\varepsilon  \right)  \prod_{i =1}^{N_1} \prod_{j =N_1+1}^{N_2} \prod_{\ell_i=1}^{k_i} \left( -m_i+m_j-\ell_i\varepsilon \right)} \,,
 \label{eq:2dkvortexplus}
   \\
&Z^{(\zeta<0)}_{2d, k}:=\lim_{\beta \to 0} Z^{(\zeta<0)}_k \nonumber \\
&=
 \sum_{ {k}_{N_1+1}+\cdots +{k}_{N_2}={k}}  \frac{(-1)^{N_1k}   \prod_{i =N_1+1}^{N_2} \prod_{j=1}^{N_{2}}
 \prod_{\ell_i=1}^{k_i}  \left( -m_i+\tilde{m}_j+ \ell_i \varepsilon \right)}
 {\prod_{i, j =N_1+1}^{N_2}  \prod_{\ell_i=1}^{k_i} \left( -m_j+m_i-(\ell_i -k_j-1)\varepsilon \right)  \prod_{i =N_1+1}^{N_2} \prod_{j =1}^{N_1}
 \prod_{\ell_i=1}^{k_i}  \left( m_i-m_j-\ell_i\varepsilon \right)} \,.
  \label{eq:2dkvortexminus}
\end{align}
}
These expressions agree with the vortex partition functions in 2d
$\mathcal{N}=(2,2)$ $U(N_1)$ gauge theory with $N_2$ fundamental and $N_2$
anti-fundamental chiral multiplets in the positive and negative FI-parameter
regions, respectively.

The 2d limit of the wall-crossing factor in \eqref{eq:WCformula1} is given by
\begin{align}
\lim_{\beta \to 0} \frac{(z_-; q)_{\infty}}{(z_+; q)_{\infty}} 
=(1-z)^{N_2-N_1+\sum_{a=1}^{N_2} \frac{\tilde{m}_a-m_a}{\varepsilon}}\,.
\label{eq:2dWCF}
\end{align}
Here we used an identity \eqref{eq:PE1} to show \eqref{eq:2dWCF}. Thus, the 2d
limit of the wall-crossing formula of the $K$-theoretic vortex partition
function leads to the relation
\begin{align}
Z^{(\zeta>0)}_{2d} =(1-z)^{N_2-N_1+\sum_{a=1}^{N_2}\frac{\tilde{m}_a-m_a}{\varepsilon}}\,Z^{(\zeta<0)}_{2d}\,,
\label{eq:2dWCformula}
\end{align}
where
\begin{align}
Z^{(\zeta)}_{2d} =1+\sum_{k=1}^{\infty}z^k Z^{(\zeta)}_{2d, k}\,.
\end{align}
This agrees with the 2d wall-crossing formula originally obtained as the
relation between vortex partition functions under the Seiberg duality between
2d $\mathcal{N}=(2,2)$ $U(N_1)$ and $U(N_2-N_1)$ gauge theories
\cite{Gomis:2014eya}, see also \cite{Benini:2012ui, Benini:2014mia}.
Specifically, the 2d wall-crossing formula \eqref{eq:2dWCformula} for $N_1=1$
and $N_2=2$ is the same as the Euler transformation of the Gauss
hypergeometric series \eqref{eq:Gauss} by setting
\begin{align}
\frac{m_1-\tilde{m}_1}{\varepsilon}=a,\,\,\,  \frac{m_1-\tilde{m}_2}{\varepsilon}=b, \,\,\, \frac{m_1-m_2 }{\varepsilon}+1=c.
\end{align}

Next we consider the 2d limit of the wall-crossing formula for 3d
$\mathcal{N}=4$ gauge theory.
{\small
\begin{align}
& Z^{(\zeta>0)}_{2d, k,(N_1,N_2)}= \lim_{\beta \to 0} Z^{(\zeta>0)}_{k,(N_1,N_2)} \nonumber \\
 &=\sum_{k_1+\cdots +k_{N_1}=k} \prod_{i,j =1}^{N_1}  \prod_{\ell_i=1}^{k_i}
  \frac{   m_i-m_j -m+(\ell_i -k_j-1)\varepsilon  } {  m_i-m_j +(\ell_i -k_j-1)\varepsilon   } \cdot
 \prod_{i =1}^{N_1} \prod_{j =N_1+1}^{N_2} \prod_{\ell_i=1}^{k_i}  \frac{-m_i+m_j-m-\ell_i\varepsilon }
 {  -m_i+m_j-\ell_i\varepsilon}   
 \label{eq:2dkvortexplusN4} \\
 & Z^{(\zeta<0)}_{2d, k,(N_1,N_2)}= \lim_{\beta \to 0}Z^{(\zeta<0)}_{k, (N_1,N_2)} \nonumber \\
 &=\sum_{k_{N_1+1}+\cdots +k_{N_2}=k} \prod_{i,j =N_1+1}^{N_2}  \prod_{\ell_i=1}^{k_i} \frac{  -m_i+m_j -m+(\ell_i -k_j-1)\varepsilon } { -m_i+m_j +(\ell_j -k_i-1)\varepsilon  } \cdot
 \prod_{i =N_1+1}^{N_2} \prod_{j =1}^{N_1} \prod_{\ell_i=1}^{k_i}  \frac{ m_i-m_j-m-\ell_i \varepsilon}
 {  m_i-m_j-\ell_i\varepsilon} 
 \label{eq:2dkvortexminusN4}
 \end{align}
}

The 2d limit of the wall-crossing factor \eqref{eq:wallcrossingN4} is given by
\begin{align}
\lim_{\beta \to 0} \prod_{s=1}^{2N_1-N_2}\frac{(z q \, t^{s+\frac{N_2-2N_1}{2}} ;q)_{\infty}}{(z \, t^{s-1+\frac{N_2-2N_1}{2}} ;q)_{\infty}} 
=(1-z)^{(N_2-2N_1)(\frac{m}{\varepsilon}-1)}
  \label{eq:2dWCFN4}
\end{align}
Thus the 2d limit of the wall-crossing formula is given by
\begin{align}
  Z^{(\zeta>0)}_{2d, (N_1,N_2)} =(1-z)^{(N_2-2N_1)(\frac{m}{\varepsilon}-1)} \,  Z^{(\zeta<0)}_{2d, (N_1,N_2)}\,,
  \label{eq:2dwallcrossingN4}
\end{align}
where
\begin{align}
Z^{(\zeta)}_{2d,(N_1,N_2)} =1+\sum_{k=1}^{\infty} z^k Z^{(\zeta)}_{2d, k,(N_1,N_2)}\,.
\end{align}
This agrees with the 2d wall-crossing formula originally derived in
\cite{Gomis:2014eya}, see also \cite{Honda:2013uca}, and is recognized as the
relation between the vortex partition functions in the Seiberg duality between
2d $\mathcal{N}=(2,2)^*$ $U(N_1)$ and $U(N_2-N_1)$ gauge theories.

Note that a 2d vortex partition function agrees with the fixed-point formula for
the integral of a cohomology class over the handsaw quiver variety.
This is analogous to a 4d instanton partition function, which agrees with an
integral of a cohomology class over the instanton moduli space.
Later, it was shown in \cite{Ohkawa:2022ebc} that the 2d wall-crossing formula
can be reinterpreted in the context of algebraic geometry and agrees with the
rational limits of the transformation formulas of Kajihara and
Halln\"as-Langmann-Noumi-Rosengren.

\section{Future directions}
\label{sec:future}

In this paper, we point out that the wall-crossing formulas of the
$K$-theoretic vortex partition functions \cite{Hwang:2017kmk} for 3d $U(N)$
supersymmetric gauge theories agree with the Euler transformations of
$q$-hypergeometric series \cite{kajihara2004euler, Halln_s_2022}. We comment on
future directions of this work.
\\
\\
\underline{Handsaw quiver of type $A_n$}
\\
Generalizations to handsaw quiver varieties of type $A_n$ correspond to the
$K$-theoretic vortex partition functions in 3d $\mathcal{N}=2$ or
$\mathcal{N}=4$ $\prod_{s=1}^{n}U(N_s)$ with $N_s < N_{s+1}$ linear quiver
supersymmetric gauge theories \cite{Bullimore:2016hdc, Hwang:2017kmk}. The
wall-crossing phenomenon for $K$-theoretic vortex partition functions
associated with the handsaw quiver varieties of type $A_n$ was partly studied
in \cite{Hwang:2017kmk}. It would be interesting to give a characterization of
the wall-crossing phenomenon studied in \cite{Hwang:2017kmk} in terms of
transformation formulas for hypergeometric series.
\\
\\
\underline{Level correspondence of $K$-theoretic $I$-functions}
\\
It has been shown \cite{Ueda:2019qhg} that a supersymmetric index on
$S^1 \times_q S^2$ factorizes into a pair of $K$-theoretic $I$-functions with
level structure introduced in \cite{ruan2019level} of the Higgs branch vacua of
a 3d theory, namely a Grassmann manifold. From the factorization into the
vortex partition functions \eqref{eq:partition1}, the $K$-theoretic vortex
partition function is closely related to the $K$-theoretic $I$-function. As
mentioned in this article, since the wall-crossing formula of the vortex
partition function is equivalent to the relation between the vortex partition
functions in a Seiberg dual pair, it is expected that the wall-crossing formula
gives a relation between $K$-theoretic $I$-functions for the Higgs branch vacua
of a 3d Seiberg dual pair.
In fact, when the moduli spaces of the vacua of a 3d Seiberg dual pair are two
Grassmannians $\mathrm{Gr}(N_1,N_2)$ and $\mathrm{Gr}(N_2-N_1,N_2)$, the
wall-crossing formula of the $k$-vortex partition function is equivalent to the
level correspondence \cite{DongWen2022}, which is the relation between the
$K$-theoretic $I$-functions with level structure for the two Grassmannians. We
plan to study the level correspondence of the $K$-theoretic $I$-functions for
manifolds realized as moduli spaces of Higgs branch vacua by using the
wall-crossing formula .

\section*{Acknowledgements}
This work is supported by Grant-in-Aid for Scientific Research 21K03382, JSPS.

\appendix
\section{$q$-Pochhammer symbol and useful identities}
\label{sec:appendix1}

The $q$-Pochhammer symbol is defined by
\begin{align}
(x;q)_{n}:=
\left\{
\begin{array}{cl}
\prod_{i=0}^{n-1}(1-x q^{i}) & (n \ge 1), \\
1 &  (n=0), \\
\frac{1}{\prod_{i=1}^{-n}(1-x q^{-i})} & (n \le -1)
\end{array}
\right.
\label{eq:qfactorial}
\end{align}
The following identities are used in the main text.
\begin{align}
&\prod_{  a =1}^{N} \frac{1 }{( q;q)_{k_a } } \cdot \prod_{1 \le   a \neq b \le N} \frac{(x_a x_b^{-1} q;q)_{k_a -k_b} }{(x_a x_b^{-1} q;q)_{k_a } }
=
\prod_{a,b=1}^N \frac{1}{(x_a x_b^{-1} q^{1+k_a-k_b};q)_{k_b}}
\label{eq:id1inapp} \\
&\prod_{1 \le   a \neq b \le N}  ( x_b x^{-1}_a q ;q )_{{k_b -k_a }}
 =(-1)^{(N-1)\sum_{a=1}^N k_a}\prod_{a=1}^{N} x _a^{N k_a -\sum_{b=1}^{N} k_b }
 \nonumber \\
& \qquad \qquad \qquad  \qquad \times q^{\frac{N}{2}(\sum_{a=1}^{N}k^2_a)-\frac{N-1}{2}\sum_{a=1}^{N} k_a-\frac{1}{2} (\sum_{a=1}^{N} k_a)^2}   \prod_{1 \le a < b \le N}
\frac{x_a q^{ k_a}-x_b q^{ k_b} }{x_a-x_b}\,,
\label{eq:id2inapp}
\\
&\sum_{k=0}^{\infty} z^k  \frac{(a;q)_k}{(q;q)_{k}} =
\frac{(a z;q)_{\infty}}{(z;q)_{\infty}} \,,
\label{eq:qbinomi} \\
&\frac{1}{(z;q)_{\infty}}=\prod_{k=1}^{\infty} \exp \left( \frac{z^k}{k} \frac{1}{1-q^k} \right)\,.
\label{eq:PE1}
\end{align}

\bibliography{refs}

\providecommand{\href}[2]{#2}\begingroup\raggedright\begin{thebibliography}{10}

\bibitem{Hwang:2017kmk}
C.~Hwang, P.~Yi, and Y.~Yoshida, ``{Fundamental Vortices, Wall-Crossing, and
  Particle-Vortex Duality},''
  \href{http://dx.doi.org/10.1007/JHEP05(2017)099}{{\em JHEP} {\bfseries 05}
  (2017) 099},
\href{http://arxiv.org/abs/1703.00213}{{\ttfamily arXiv:1703.00213 [hep-th]}}.

\bibitem{kajihara2004euler}
Y.~Kajihara, ``Euler transformation formula for multiple basic hypergeometric
  series of type a and some applications,'' {\em Advances in Mathematics}
  {\bfseries 187} no.~1, (2004) 53--97.

\bibitem{Halln_s_2022}
M.~Hallnäs, E.~Langmann, M.~Noumi, and H.~Rosengren, ``Higher order deformed
  elliptic ruijsenaars operators,''
  \href{http://dx.doi.org/10.1007/s00220-022-04360-7}{{\em Commun. Math. Phys.}
  {\bfseries 392} no.~2, (Apr., 2022) 659–689},
  \href{http://arxiv.org/abs/arXiv:2105.02536}{{\ttfamily
  arXiv:arXiv:2105.02536 [math-ph]}}.

\bibitem{Romelsberger:2005eg}
C.~Romelsberger, ``{Counting chiral primaries in N = 1, d=4 superconformal
  field theories},''
  \href{http://dx.doi.org/10.1016/j.nuclphysb.2006.03.037}{{\em Nucl. Phys. B}
  {\bfseries 747} (2006) 329--353},
  \href{http://arxiv.org/abs/hep-th/0510060}{{\ttfamily arXiv:hep-th/0510060}}.

\bibitem{Kinney:2005ej}
J.~Kinney, J.~M. Maldacena, S.~Minwalla, and S.~Raju, ``{An Index for 4
  dimensional super conformal theories},''
  \href{http://dx.doi.org/10.1007/s00220-007-0258-7}{{\em Commun. Math. Phys.}
  {\bfseries 275} (2007) 209--254},
  \href{http://arxiv.org/abs/hep-th/0510251}{{\ttfamily arXiv:hep-th/0510251}}.

\bibitem{Kim:2009wb}
S.~Kim, ``{The Complete superconformal index for N=6 Chern-Simons theory},''
  \href{http://dx.doi.org/10.1016/j.nuclphysb.2009.06.025}{{\em Nucl. Phys. B}
  {\bfseries 821} (2009) 241--284},
  \href{http://arxiv.org/abs/0903.4172}{{\ttfamily arXiv:0903.4172 [hep-th]}}.
  [Erratum: Nucl.Phys.B 864, 884 (2012)].

\bibitem{Hama:2011ea}
N.~Hama, K.~Hosomichi, and S.~Lee, ``{SUSY Gauge Theories on Squashed
  Three-Spheres},'' \href{http://dx.doi.org/10.1007/JHEP05(2011)014}{{\em JHEP}
  {\bfseries 05} (2011) 014}, \href{http://arxiv.org/abs/1102.4716}{{\ttfamily
  arXiv:1102.4716 [hep-th]}}.

\bibitem{Imamura:2011su}
Y.~Imamura and S.~Yokoyama, ``{Index for three dimensional superconformal field
  theories with general R-charge assignments},''
  \href{http://dx.doi.org/10.1007/JHEP04(2011)007}{{\em JHEP} {\bfseries 04}
  (2011) 007}, \href{http://arxiv.org/abs/1101.0557}{{\ttfamily arXiv:1101.0557
  [hep-th]}}.

\bibitem{Benini:2015noa}
F.~Benini and A.~Zaffaroni, ``{A topologically twisted index for
  three-dimensional supersymmetric theories},''
  \href{http://dx.doi.org/10.1007/JHEP07(2015)127}{{\em JHEP} {\bfseries 07}
  (2015) 127},
\href{http://arxiv.org/abs/1504.03698}{{\ttfamily arXiv:1504.03698 [hep-th]}}.

\bibitem{Pasquetti:2011fj}
S.~Pasquetti, ``{Factorisation of N = 2 Theories on the Squashed 3-Sphere},''
  \href{http://dx.doi.org/10.1007/JHEP04(2012)120}{{\em JHEP} {\bfseries 04}
  (2012) 120}, \href{http://arxiv.org/abs/1111.6905}{{\ttfamily arXiv:1111.6905
  [hep-th]}}.

\bibitem{Hwang:2012jh}
C.~Hwang, H.-C. Kim, and J.~Park, ``{Factorization of the 3d superconformal
  index},'' \href{http://dx.doi.org/10.1007/JHEP08(2014)018}{{\em JHEP}
  {\bfseries 08} (2014) 018}, \href{http://arxiv.org/abs/1211.6023}{{\ttfamily
  arXiv:1211.6023 [hep-th]}}.

\bibitem{Beem:2012mb}
C.~Beem, T.~Dimofte, and S.~Pasquetti, ``{Holomorphic Blocks in Three
  Dimensions},'' \href{http://dx.doi.org/10.1007/JHEP12(2014)177}{{\em JHEP}
  {\bfseries 12} (2014) 177},
\href{http://arxiv.org/abs/1211.1986}{{\ttfamily arXiv:1211.1986 [hep-th]}}.

\bibitem{Taki:2013opa}
M.~Taki, ``{Holomorphic Blocks for 3d Non-abelian Partition Functions},''
  \href{http://arxiv.org/abs/1303.5915}{{\ttfamily arXiv:1303.5915 [hep-th]}}.

\bibitem{Fujitsuka:2013fga}
M.~Fujitsuka, M.~Honda, and Y.~Yoshida, ``{Higgs branch localization of 3d N= 2
  theories},'' \href{http://dx.doi.org/10.1093/ptep/ptu158}{{\em PTEP}
  {\bfseries 2014} no.~12, (2014) 123B02},
\href{http://arxiv.org/abs/1312.3627}{{\ttfamily arXiv:1312.3627 [hep-th]}}.

\bibitem{Benini:2013yva}
F.~Benini and W.~Peelaers, ``{Higgs branch localization in three dimensions},''
  \href{http://dx.doi.org/10.1007/JHEP05(2014)030}{{\em JHEP} {\bfseries 05}
  (2014) 030},
\href{http://arxiv.org/abs/1312.6078}{{\ttfamily arXiv:1312.6078 [hep-th]}}.

\bibitem{Hwang:2015wna}
C.~Hwang and J.~Park, ``{Factorization of the 3d superconformal index with an
  adjoint matter},'' \href{http://dx.doi.org/10.1007/JHEP11(2015)028}{{\em
  JHEP} {\bfseries 11} (2015) 028},
  \href{http://arxiv.org/abs/1506.03951}{{\ttfamily arXiv:1506.03951
  [hep-th]}}.

\bibitem{Hanany:2003hp}
A.~Hanany and D.~Tong, ``{Vortices, instantons and branes},''
  \href{http://dx.doi.org/10.1088/1126-6708/2003/07/037}{{\em JHEP} {\bfseries
  07} (2003) 037}, \href{http://arxiv.org/abs/hep-th/0306150}{{\ttfamily
  arXiv:hep-th/0306150}}.

\bibitem{Eto:2005yh}
M.~Eto, Y.~Isozumi, M.~Nitta, K.~Ohashi, and N.~Sakai, ``{Moduli space of
  non-Abelian vortices},''
  \href{http://dx.doi.org/10.1103/PhysRevLett.96.161601}{{\em Phys. Rev. Lett.}
  {\bfseries 96} (2006) 161601},
  \href{http://arxiv.org/abs/hep-th/0511088}{{\ttfamily arXiv:hep-th/0511088}}.

\bibitem{Eto:2006pg}
M.~Eto, Y.~Isozumi, M.~Nitta, K.~Ohashi, and N.~Sakai, ``{Solitons in the Higgs
  phase: The Moduli matrix approach},''
  \href{http://dx.doi.org/10.1088/0305-4470/39/26/R01}{{\em J. Phys. A}
  {\bfseries 39} (2006) R315--R392},
  \href{http://arxiv.org/abs/hep-th/0602170}{{\ttfamily arXiv:hep-th/0602170}}.

\bibitem{Nekrasov:2004vw}
N.~Nekrasov and S.~Shadchin, ``{ABCD of instantons},''
  \href{http://dx.doi.org/10.1007/s00220-004-1189-1}{{\em Commun. Math. Phys.}
  {\bfseries 252} (2004) 359--391},
\href{http://arxiv.org/abs/hep-th/0404225}{{\ttfamily arXiv:hep-th/0404225
  [hep-th]}}.

\bibitem{Hwang:2014uwa}
C.~Hwang, J.~Kim, S.~Kim, and J.~Park, ``{General instanton counting and 5d
  SCFT},'' \href{http://dx.doi.org/10.1007/JHEP07(2015)063,
  10.1007/JHEP04(2016)094}{{\em JHEP} {\bfseries 07} (2015) 063},
  \href{http://arxiv.org/abs/1406.6793}{{\ttfamily arXiv:1406.6793 [hep-th]}}.
[Addendum: JHEP04,094(2016)].

\bibitem{Hori:2014tda}
K.~Hori, H.~Kim, and P.~Yi, ``{Witten Index and Wall Crossing},''
\href{http://arxiv.org/abs/1407.2567}{{\ttfamily arXiv:1407.2567 [hep-th]}}.

\bibitem{KajiharaNoumi2003}
Y.~Kajihara and M.~Noumi, ``Multiple elliptic hypergeometric series. an
  approach from the cauchy determinant,'' {\em Indagationes Mathematicae}
  {\bfseries 14} no.~3-4, (2003) 395--421,
  \href{http://arxiv.org/abs/arXiv:math/0306219}{{\ttfamily
  arXiv:arXiv:math/0306219 [math.CA]}}.

\bibitem{Ohkawa:2022ebc}
R.~Ohkawa and Y.~Yoshida, ``{Wall-crossing for vortex partition function and
  handsaw quiver variety},''
  \href{http://dx.doi.org/10.1016/j.geomphys.2023.104904}{{\em J. Geom. Phys.}
  {\bfseries 191} (2023) 104904},
  \href{http://arxiv.org/abs/2208.00435}{{\ttfamily arXiv:2208.00435
  [math.AG]}}.

\bibitem{Yoshida:2011au}
Y.~Yoshida, ``{Localization of Vortex Partition Functions in $\mathcal{N}=(2,2)
  $ Super Yang-Mills theory},''
  \href{http://arxiv.org/abs/1101.0872}{{\ttfamily arXiv:1101.0872 [hep-th]}}.

\bibitem{Bonelli:2011fq}
G.~Bonelli, A.~Tanzini, and J.~Zhao, ``{Vertices, Vortices and Interacting
  Surface Operators},'' \href{http://dx.doi.org/10.1007/JHEP06(2012)178}{{\em
  JHEP} {\bfseries 06} (2012) 178},
  \href{http://arxiv.org/abs/1102.0184}{{\ttfamily arXiv:1102.0184 [hep-th]}}.

\bibitem{Hollowood:2003cv}
T.~J. Hollowood, A.~Iqbal, and C.~Vafa, ``{Matrix models, geometric engineering
  and elliptic genera},''
  \href{http://dx.doi.org/10.1088/1126-6708/2008/03/069}{{\em JHEP} {\bfseries
  03} (2008) 069}, \href{http://arxiv.org/abs/hep-th/0310272}{{\ttfamily
  arXiv:hep-th/0310272}}.

\bibitem{Benini:2013nda}
F.~Benini, R.~Eager, K.~Hori, and Y.~Tachikawa, ``{Elliptic genera of
  two-dimensional N=2 gauge theories with rank-one gauge groups},''
  \href{http://dx.doi.org/10.1007/s11005-013-0673-y}{{\em Lett. Math. Phys.}
  {\bfseries 104} (2014) 465--493},
\href{http://arxiv.org/abs/1305.0533}{{\ttfamily arXiv:1305.0533 [hep-th]}}.

\bibitem{Benini:2013xpa}
F.~Benini, R.~Eager, K.~Hori, and Y.~Tachikawa, ``{Elliptic Genera of 2d
  ${\mathcal{N}}$ = 2 Gauge Theories},''
  \href{http://dx.doi.org/10.1007/s00220-014-2210-y}{{\em Commun. Math. Phys.}
  {\bfseries 333} no.~3, (2015) 1241--1286},
\href{http://arxiv.org/abs/1308.4896}{{\ttfamily arXiv:1308.4896 [hep-th]}}.

\bibitem{Gomis:2014eya}
J.~Gomis and B.~Le~Floch, ``{M2-brane surface operators and gauge theory
  dualities in Toda},'' \href{http://dx.doi.org/10.1007/JHEP04(2016)183}{{\em
  JHEP} {\bfseries 04} (2016) 183},
\href{http://arxiv.org/abs/1407.1852}{{\ttfamily arXiv:1407.1852 [hep-th]}}.

\bibitem{Benini:2012ui}
F.~Benini and S.~Cremonesi, ``{Partition Functions of ${\mathcal{N}=(2,2)}$
  Gauge Theories on S$^{2}$ and Vortices},''
  \href{http://dx.doi.org/10.1007/s00220-014-2112-z}{{\em Commun. Math. Phys.}
  {\bfseries 334} no.~3, (2015) 1483--1527},
\href{http://arxiv.org/abs/1206.2356}{{\ttfamily arXiv:1206.2356 [hep-th]}}.

\bibitem{Benini:2014mia}
F.~Benini, D.~S. Park, and P.~Zhao, ``{Cluster Algebras from Dualities of 2d
  ${\mathcal{N}}$ = (2, 2) Quiver Gauge Theories},''
  \href{http://dx.doi.org/10.1007/s00220-015-2452-3}{{\em Commun. Math. Phys.}
  {\bfseries 340} (2015) 47--104},
  \href{http://arxiv.org/abs/1406.2699}{{\ttfamily arXiv:1406.2699 [hep-th]}}.

\bibitem{Honda:2013uca}
D.~Honda and T.~Okuda, ``{Exact results for boundaries and domain walls in 2d
  supersymmetric theories},''
  \href{http://dx.doi.org/10.1007/JHEP09(2015)140}{{\em JHEP} {\bfseries 09}
  (2015) 140},
\href{http://arxiv.org/abs/1308.2217}{{\ttfamily arXiv:1308.2217 [hep-th]}}.

\bibitem{Bullimore:2016hdc}
M.~Bullimore, T.~Dimofte, D.~Gaiotto, J.~Hilburn, and H.-C. Kim, ``{Vortices
  and Vermas},'' \href{http://dx.doi.org/10.4310/ATMP.2018.v22.n4.a1}{{\em Adv.
  Theor. Math. Phys.} {\bfseries 22} (2018) 803--917},
  \href{http://arxiv.org/abs/1609.04406}{{\ttfamily arXiv:1609.04406
  [hep-th]}}.

\bibitem{Ueda:2019qhg}
K.~Ueda and Y.~Yoshida, ``{3d $ \mathcal{N} $ = 2 Chern-Simons-matter theory,
  Bethe ansatz, and quantum $K$-theory of Grassmannians},''
  \href{http://dx.doi.org/10.1007/JHEP08(2020)157}{{\em JHEP} {\bfseries 08}
  (2020) 157}, \href{http://arxiv.org/abs/1912.03792}{{\ttfamily
  arXiv:1912.03792 [hep-th]}}.

\bibitem{ruan2019level}
Y.~Ruan and M.~Zhang, ``{The level structure in quantum K-theory and mock theta
  functions},'' \href{http://arxiv.org/abs/1804.06552}{{\ttfamily
  arXiv:1804.06552 [math.AG]}}.

\bibitem{DongWen2022}
H.~Dong and Y.~Wen, ``{Level correspondence of the $K$-theoretic $I$-function
  in Grassmann duality},'' \href{http://dx.doi.org/10.1017/fms.2022.28}{{\em
  Forum of Mathematics, Sigma} {\bfseries 10} (2022) e44},
  \href{http://arxiv.org/abs/arXiv:2004.10661}{{\ttfamily
  arXiv:arXiv:2004.10661 [math.AG]}}.


\end{thebibliography}\endgroup

\end{document}